\documentclass[twocolumn,twocolappendix]{aastex631}

\usepackage{graphicx}	
\usepackage{amsmath}	
\usepackage{amssymb}	

\usepackage{natbib}
\usepackage{enumitem}
\usepackage[flushleft]{threeparttable}
\usepackage{colortbl}
\usepackage[normalem]{ulem}
\usepackage{color}
\usepackage{bm} 
\usepackage{gensymb}
\definecolor{darkblue}{rgb}{0.0,0.0,0.8}
\definecolor{darkred}{rgb}{0.75,0.,0.25}
\definecolor{darkorange}{rgb}{1,0.3,0.0}
\definecolor{darkgreen}{rgb}{0.0,0.6,0.0}
\definecolor{darkpurple}{rgb}{0.8,0.,0.9}
\definecolor{brown}{rgb}{0.65,.16,0.16}
\definecolor{grey}{rgb}{0.4,0.5,0.6}
\definecolor{white}{rgb}{1,1,1}
\definecolor{trolleygrey}{rgb}{0.5, 0.5, 0.5}
\definecolor{lavender}{rgb}{0.835,0.812,0.969}
\definecolor{pastelorange}{rgb}{0.99,0.92,0.82}
\definecolor{pastelblue}{rgb}{0.85,0.93,0.99}

\newcommand{\jampy}{{\sc JamPy}}
\newcommand{\scf}{{\sc scalefree}}
\newcommand{\agama}{{\sc Agama}}

\newcommand{\gaia}{{\sl Gaia}}
\newcommand{\hst}{{\sl HST}}
\newcommand{\jwst}{{\sl JWST}}

\newcommand{\msun}{\rm M_\odot}
\newcommand{\kms}{\rm km\,s^{-1}}
\newcommand{\masyr}{\rm mas\,yr^{-1}}

\begin{document}

\title{HSTPROMO Internal Proper Motion Kinematics of Dwarf Spheroidal Galaxies:\\ II.~Velocity Anisotropy and Dark Matter Cusp Slope of Sculptor}

\correspondingauthor{Eduardo Vitral}
\email{eduardo.vitral@roe.ac.uk}

\author[0000-0002-2732-9717]{Eduardo Vitral}\thanks{Royal Society Newton International Fellow}
\affiliation{Institute for Astronomy, University of Edinburgh, Royal Observatory, Blackford Hill, Edinburgh EH9 3HJ, UK}
\affiliation{Space Telescope Science Institute, 3700 San Martin Drive, Baltimore, MD 21218, USA}

\author[0000-0001-7827-7825]{Roeland P. van der Marel}
\affiliation{Space Telescope Science Institute, 3700 San Martin Drive, Baltimore, MD 21218, USA}
\affiliation{Center for Astrophysical Sciences, The William H. Miller III Department of Physics \& Astronomy, Johns Hopkins University, Baltimore, MD 21218, USA}

\author[0000-0001-8368-0221]{Sangmo Tony Sohn}
\affiliation{Space Telescope Science Institute, 3700 San Martin Drive, Baltimore, MD 21218, USA}
\affiliation{Dept. of Astronomy \& Space Science, Kyung Hee University, Gyeonggi-do 17104, Republic of Korea}

\author{Jorge Pe\~narrubia}
\affiliation{Institute for Astronomy, University of Edinburgh, Royal Observatory, Blackford Hill, Edinburgh EH9 3HJ, UK}
\affiliation{Centre for Statistics, University of Edinburgh, School of Mathematics, Edinburgh EH9 3FD, UK}

\author[0000-0002-9820-1219]{Ekta~Patel}\thanks{NASA Hubble Fellow}
\affiliation{Department of Physics and Astronomy, University of Utah, 115 South 1400 East, Salt Lake City, Utah 84112, USA}
\affiliation{Department of Astrophysics and Planetary Sciences, Villanova University, 800 E. Lancaster Ave, Villanova, PA 19085, USA}

\author[0000-0002-1343-134X]{Laura L. Watkins}
\affiliation{AURA for the European Space Agency (ESA), Space Telescope Science Institute, 3700 San Martin Drive, Baltimore, MD 21218, USA}

\author[0000-0001-9673-7397]{Mattia Libralato}
\affiliation{INAF - Osservatorio Astronomico di Padova, Vicolo dell'Osservatorio 5, Padova I-35122, Italy}

\author[0000-0001-7494-5910]{Kevin A. McKinnon}
\affiliation{David A. Dunlap Department of Astronomy \& Astrophysics, University of Toronto, 50 St. George Street, Toronto, ON M5S 3H4, Canada}

\author[0000-0003-3858-637X]{Andrea Bellini}
\affiliation{Space Telescope Science Institute, 3700 San Martin Drive, Baltimore, MD 21218, USA}

\author[0000-0003-4922-5131]{Andrés del Pino}
\affiliation{Instituto de Astrofísica de Andalucía, CSIC, Glorieta de la Astronomía, 18080 Granada, Spain}

\author[0000-0001-8354-7279]{Paul Bennet}
\affiliation{Space Telescope Science Institute, 3700 San Martin Drive, Baltimore, MD 21218, USA}


\begin{abstract}
We analyze three epochs of \hst\ imaging over 20 years for the Sculptor dwarf spheroidal galaxy, measuring precise proper motions for 119 stars and combining them with 1,760 existing line-of-sight velocities. This catalog yields the first radially-resolved 3D velocity dispersion profiles for Sculptor. We confirm mild oblate rotation, with major-axis velocities reaching $\sim 2~\kms$ beyond $20\farcm0$.
Using a methodology similar to that in the first paper in this series, we solve the Jeans equations in oblate axisymmetric geometry to infer the galaxy’s mass profile. Our modeling reveals a significant degeneracy due to the unknown galaxy inclination, which is overlooked under spherical symmetry assumptions. This degeneracy allows acceptable fits across a range of dark matter profiles, from cuspy to cored. While we do not directly constrain the inclination with our Jeans models, higher-order line-of-sight velocity moments provide useful additional constraints: comparisons with \scf\ models from \cite{deBruijne+96} favor highly flattened (more face-on) configurations. Adopting an inclination well consistent with these comparisons ($i = 57\fdg1$), we find, alongside radial velocity anisotropy, a dark matter density slope of $\Gamma_{\rm dark} = 0.29^{+ 0.31}_{- 0.41}$ within the radial extent of the 3D velocity data (i.e. within $\sim 120 - 240$~pc), ruling out a cusp with $\Gamma_{\rm dark} \leq -1$ at 99.8\% confidence. This confidence increases for lower inclinations and decreases drastically for nearly edge-on configurations. The results qualitatively agree with $\Lambda$CDM, SIDM, and Fuzzy DM scenarios that predict core formation, while our specific measurements provide quantitative constraints on the prescriptions of feedback, cross sections, or particle masses required by these models, respectively.
\end{abstract}


\section{Introduction} \label{sec: intro}

The nature of dark matter (DM) remains one of the most profound open questions in modern astrophysics, with far-reaching implications for both cosmology and our understanding of galaxy formation \citep*{Bertone+05,Silk&Mamon12}. Among the environments in which DM can be studied, dwarf spheroidal galaxies (dSphs) stand out as particularly valuable laboratories \citep{Strigari+13, Collins&Read22}. These small, faint systems are among the most DM-dominated objects known \citep{Pryor&Kormendy&Kormendy90}, making them highly sensitive to the properties of DM on small scales \citep{Bullock&Boylan-Kolchin17}. Their relatively simple stellar populations and quiescent evolutionary histories (\citealt*{Tolstoy+09}; \citealt{Savino+25}) also provide a cleaner view of internal dynamical processes than the more complex environments found in larger galaxies \citep{Grebel+09}.

Accurately inferring the internal kinematics of dSphs remains a long-standing challenge. Analyses based solely on line-of-sight (LOS) velocities\footnote{Motions toward or away from the observer, measured through spectroscopic data typically obtained from ground-based observatories.} are fundamentally limited by the well-known degeneracy between the mass distribution and the orbital anisotropy of the stellar component \citep{Binney&Mamon82}. This degeneracy hampers efforts to robustly distinguish between different DM density profiles, complicating tests of predictions from $\Lambda$ Cold Dark Matter ($\Lambda$CDM) and alternative cosmological models (see \citealt{Salucci19} and \citealt{Boldrini21} for reviews and references). Overcoming this limitation typically requires either modeling techniques that incorporate higher-order velocity moments \citep*{Chaname+08,Read&Steger17} -- which remain difficult to constrain with current LOS data -- or the inclusion of transverse stellar motions in the plane of the sky \citep{vanderMarel&Anderson&Anderson10}. The latter has historically been inaccessible for galaxies beyond the Milky Way \citep{GaiaCollaborationHelmi+18}, owing to the simultaneous need for long time baselines (typically $\gtrsim 10$~yrs) and high-precision astrometry to reach sufficiently small uncertainties.

The advent of space-based astrometry has now opened new avenues for addressing this limitation. Transverse velocity measurements, or proper motions (PMs), can be obtained with high precision and, when combined with LOS velocities, enable the reconstruction of full three-dimensional velocity vectors for stars in nearby dSphs \citep{Massari+17, Massari+20, delPino+22, Libralato+23, Vitral+24, McKinnon+24}. These additional kinematic dimensions significantly enhance the constraining power of dynamical models and allow for a more complete and direct view of a system’s internal dynamics than can be obtained from LOS velocities alone \citep{Read+21}.

In a series of papers beginning with \citeauthor{Vitral+24}~(\citeyear{Vitral+24}, hereafter Paper~I), we aim to construct comprehensive PM catalogs for nearby dSphs and use them, alongside LOS velocities, to build detailed dynamical models of these systems. By systematically applying these techniques, we seek to refine our understanding of their underlying DM distributions and evaluate the influence of baryonic processes in shaping both their internal kinematics and DM density profiles. In the first paper of this series, we analyzed the Draco dSph and found that its internal dynamics are consistent with a cuspy DM profile, as predicted by DM-only $\Lambda$CDM simulations \citep*{Navarro+97}. In this second study, we turn our attention to the Sculptor dSph (also known as PGC3589, MCG-06-03-015, or ESO351-30), a well-studied system \citep{Battaglia+08,Lokas09, Walker&Penarrubia&Penarrubia11,Richardson&Fairbairn14, Massari+17,Arroyo-Polonio+25} which, until now, lacked a large and homogeneous PM dataset with uncertainties smaller than its intrinsic velocity dispersion. Such precision is essential to mitigate the impact of measurement errors and systematic biases in dynamical mass estimates \citep{Watkins+13,Vitral+22}, as discussed in Paper~I. Among Sculptor’s notable features is its higher stellar mass relative to Draco \citep{Martin+08, deBoer+12}, placing the two galaxies on opposite sides of the threshold where baryonic processes are thought to flatten DM cusps \citep{Fitts+17}, despite their broadly similar structural properties \citep{Munoz+18}.

With this background in mind, the remainder of the paper is structured as follows. Section~\ref{sec: data} outlines the general properties of Sculptor relevant to our modeling and describes the kinematic datasets used throughout the study. Section~\ref{sec: methods} details the modeling techniques employed to analyze the galaxy. Sections~\ref{sec: results} and \ref{sec: robust} present our main results and assess their robustness, respectively. Section~\ref{sec: discussion} discusses the implications of our findings for mass modeling and broader cosmological questions. Finally, Section~\ref{sec: conclusion} summarizes our key conclusions.

\section{Sculptor Data and General Characteristics} \label{sec: data}

\begin{deluxetable*}{llllllll}
\tablecaption{Overview of Sculptor structural parameters.}
\label{tab: overview}
\tablewidth{750pt}
\renewcommand{\arraystretch}{1.1}
\tabcolsep=3.2pt
\tabletypesize{\scriptsize}
\tablehead{
\colhead{Reference} &
\colhead{Data} &
\colhead{$\alpha_{0}$} &
\colhead{$\delta_{0}$} &
\colhead{$\theta$} &
\colhead{$\epsilon$} &
\colhead{$R_{h}$} \\
\colhead{(1)} &
\colhead{(2)} &
\colhead{(3)} &
\colhead{(4)} &
\colhead{(5)} &
\colhead{(6)} &
\colhead{(7)}
}
\startdata
This work & \gaia~EDR3 &$ 15\fdg02328 \pm 8\farcs2$ & $-33\fdg71504 \pm 4\farcs8$ & $95\fdg4 \pm 1\fdg6$ & $0.272 \pm 0.013$ & $11\farcm12 \pm 0\farcm17$ \\
\protect\cite{Munoz+18} & MegaCam & $15\fdg01830 \pm 0\farcs3$ & $ -33\fdg71860 \pm 2\farcs6$ & $92\fdg0 \pm 1\fdg0$ & $0.330 \pm 0.010$ & $11\farcm17 \pm 0\farcm05$ & \\
\enddata
\begin{tablenotes}
\scriptsize
\item \textsc{Notes} --  
Columns are 
\textbf{(1)} Reference where the values are reported; 
\textbf{(2)} data source of respective estimates (columns 3--7); 
\textbf{(3)} right ascension of Sculptor center; 
\textbf{(4)} declination of Sculptor center; 
\textbf{(5)} Major axis projected angle in the sky, from North to East; 
\textbf{(6)} projected ellipticity in the sky, defined as $1 - b/a$, with $a$ and $b$ the major and minor axes of the projected ellipse, respectively; 
\textbf{(7)} 2D half-number radius of a Plummer model fit (i.e. the major axis of the ellipse that contains half the stellar count).
\end{tablenotes}
\end{deluxetable*}

\begin{figure}
\centering
\includegraphics[width=\hsize]{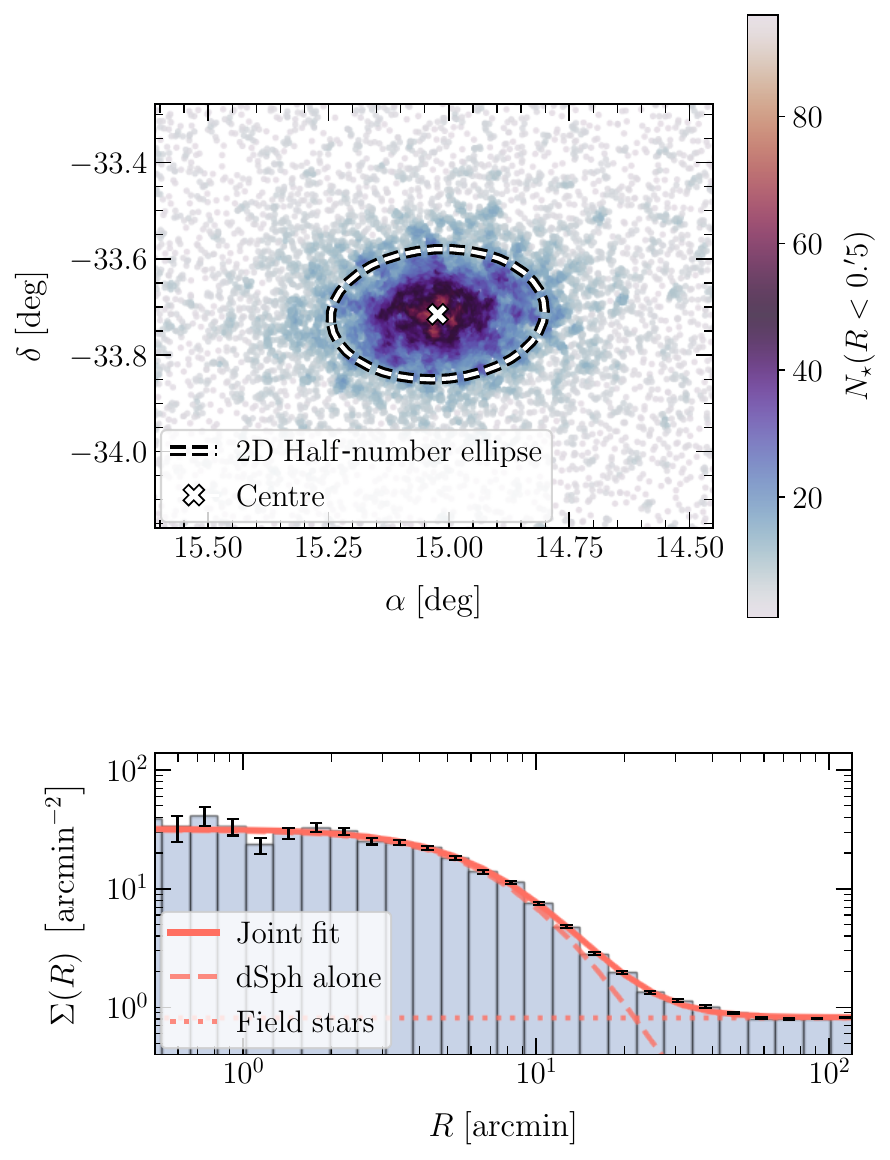}
\caption{\textit{Geometry and density:} The \textbf{top} panel presents a 
sky view of Sculptor, with \gaia~EDR3 measurements scattered 
and color-coded by the number of neighbors within a $0\farcm5$ 
projected radius. The 2D half-number ellipse and the center from our 
fits are shown as a white dashed line and a white cross, respectively. 
The \textbf{lower} panel shows the observed surface number density of stars within an annulus at radius $R$ -- the projected distance from the center of Sculptor -- in blue, along with the corresponding model fit, decomposed into its individual components: the dSph alone in dashed red, the interlopers in dotted 
red, and their sum in solid red. The figure highlights a satisfactory 
fit of Sculptor's structural parameters to the \gaia~EDR3 data.}
\label{fig: structural-params}
\end{figure}

\subsection{Structural Parameters} \label{ssec: density}

We begin by defining and presenting the structural parameters used as input throughout this work -- namely, the shape of the stellar mass density profile, characterized by the total stellar mass, its scale radius, and the position of the galaxy’s center of mass (hereafter simply center). The position of Sculptor's center, as listed in the \cite{McConnachie12} 
catalog and the Local Volume Database \citep{Pace+24}, originates from 
its initial discovery by \cite{Shapley38}. Recently, 
\cite{Munoz+18} refined Sculptor's center and structural parameters 
using Canada France Hawaii Telescope's MegaCam imaging. Their analysis provided more 
reliable estimates based on improved stellar measurements. In this
study, we adopt a similar fitting methodology to \cite{Munoz+18}, 
also employed in Paper~I (see the latter's section~2.1.1 and 
eq.~A2). Specifically, we fit an axisymmetric \cite{Plummer1911} model 
to \gaia~EDR3 star counts, incorporating a constant-density interloper 
background to derive Sculptor's structural parameters and central 
position.

Our goodness-of-fit results are summarized in Figure \ref{fig: structural-params}, which compares the Gaia EDR3 data with our best-fitting structural model.
The top panel shows a centered sky view of Sculptor, where individual Gaia sources are color-coded by the local stellar count within $0\farcm5$. The fitted center (white cross) lies close to the observed density maximum, and the model’s 2D half-number ellipse (white dashed line) traces the projected stellar distribution well.
The lower panel presents the azimuthally averaged surface number density as a function of projected radius $R$. The observed profile (blue) is well matched by the sum (solid red) of two components: the dSph itself (dashed red) and a uniform interloper background (dotted red).
The good agreement in both the spatial distribution and radial profile indicates that our model captures Sculptor’s sky orientation, central position, and overall surface density structure. The parameters derived from this fit 
are presented in Table~\ref{tab: overview}, alongside those reported by 
\cite{Munoz+18}. 

The overall shape of Sculptor -- quantified here by its position angle $\theta$, ellipticity $\epsilon$, and projected half-number radius $R_{h}$ -- agrees closely with the results of \cite{Munoz+18}. In our convention, $\epsilon \equiv 1 - b/a$, where $a$ and $b$ are the major and minor axes of the best-fitting photometric ellipse, and $\theta$ is the position angle of the major axis measured from north toward east. The half-number radius $R_{h}$ corresponds to the major axis length of the ellipse enclosing half the stars in a Plummer model fit. Our fitted center lies in between the \cite{Shapley38} and \cite{Munoz+18} values, with respective distances of $52\farcs5$ and $19\farcs7$ from each.

\begin{figure}
\centering
\includegraphics[width=\hsize]{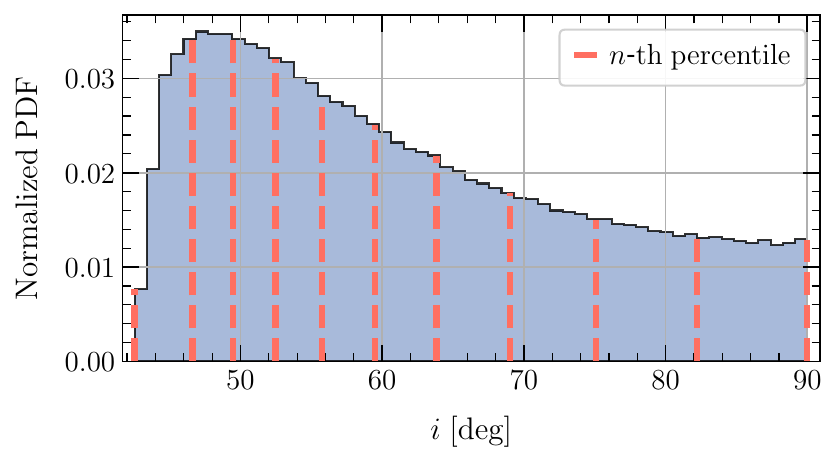}
\caption{\textit{Inclinations:} Probability distribution function of inclinations for Sculptor, computed using the formalism described in Paper~I (section~2.1.3). 
The \textit{dashed red} lines represent the $n$-th percentiles of the 
distribution, where $n$ ranges from zero to a hundred, in intervals 
of ten.}
\label{fig: inc-dist}
\end{figure}

The flattening of Sculptor closely resembles that of Draco, resulting 
in a probability distribution function (PDF) of possible inclinations 
that follows a similarly shaped curve. This PDF is calculated using the 
formalism from Paper~I (see section~2.1.3), which is based on the statistical distribution of elliptical galaxy shapes inferred by \cite{Lambas+92}. The resulting distribution is shown in Figure~\ref{fig: inc-dist}. The $[16, \, 50, \, 84]$ 
percentiles for the inclination and 3D flattening distributions are 
$[48\fdg34, \, 59\fdg50, \, 77\fdg82]$ and $[0.38, \, 
0.60, \, 0.71]$, respectively.

Finally, we consider Sculptor's stellar mass as estimated by 
\cite{deBoer+12} based on its star formation history and 
color-magnitude diagram (CMD). They report a stellar mass of 
$7.8 \times 10^{6}~\msun$ within a $1\fdg0$ elliptical
radius.\footnote{Their definition of an elliptical radius, clarified through 
private communication with T. J. L. de Boer, involves projecting the 
($\alpha$, $\delta$) data onto ($x$, $y$) Cartesian coordinates, 
rotating them by Sculptor's position angle, and dividing the $y$ 
coordinate by $(1 - \epsilon)$. The elliptical radius is then 
calculated as $\sqrt{x'^{2} + y'^{2}}$, where $x'$ and $y'$ are the 
rotated and scaled Cartesian coordinates.} Extrapolating this measurement 
to infinity, assuming an axisymmetric Plummer density profile, yields a 
total stellar mass of $8.08 \times 10^{6}~\msun$, which is adopted 
throughout this work.\footnote{For context, this value relates to a DM halo mass of $M_{\rm dark} = 4.85 \times 10^{9}~\msun$ in $\Lambda$CDM cosmology, using the relation for low-mass dwarfs from \cite{Read+17}.}

\subsection{Line-of-Sight Velocities} \label{ssec: los}

Similarly to Paper~I, we use the spectroscopic catalog from 
\cite{Walker+23} to obtain LOS velocities for our mass 
modeling. Following the same data cleaning procedures detailed in 
Paper~I (see section~2.2.1), we identify 1151 Sculptor members 
from this catalog. To augment this dataset, we incorporate the recently 
published spectroscopic measurements by \cite{Tolstoy+23}, which are 
derived from \gaia\ and VLT/FLAMES observations. We clean the 
\cite{Tolstoy+23} catalog following their recommended criteria: setting 
the $z$ coefficient\footnote{\cite{Tolstoy+23} define $z$ as a 
membership score based on \gaia's PMs and parallaxes.} 
$\leq 14.2$, enforcing ${\rm S/N} > 20$, and selecting only stars 
classified as members in their analysis. This procedure yields an 
additional 1123 LOS velocities. For both catalogs, we apply corrections 
for perspective motions using the formalism from \citeauthor{vanderMarel+02}~(\citeyear{vanderMarel+02}, eq.~13).

To leverage the strengths and coverage of both datasets, we merge them 
using the following approach: 

\begin{enumerate}
    \item We first identify common stars in the 
    two catalogs by performing a symmetric sky match with a maximum 
    separation of $1\farcs0$. This matched subset serves as the basis for 
    comparison. For each catalog, we compute zero-points for LOS velocities 
    using the following steps:
    \begin{enumerate}
        \item Compute the weighted mean of the LOS velocities for the 
        matched stars in each catalog.
        \item For each catalog, calculate the distribution of velocity 
        offsets as the difference between the LOS velocities of individual 
        stars and the weighted mean of the matched subset.
        \item Derive the zero-point correction for each catalog by 
        taking the weighted mean of these velocity offset distributions.
    \end{enumerate}
    \item The computed zero-point corrections are applied to the LOS 
    velocities in each catalog,\footnote{This common treatment is done since we lack an independent means of determining which catalog’s zero-point is intrinsically more accurate. Instead, we align both to a common scale using uncertainty-weighted averages, which incorporate each catalog’s reported precision. The resulting shifts are negligible compared to the velocity scales of interest in this work (e.g., $\sim 2~\kms$ rotation and $\sim 10~\kms$ dispersion): $0.03 \pm 0.01~\kms$ for the \cite{Walker+23} data and $-0.32 \pm 0.03~\kms$ for the \cite{Tolstoy+23} data.} ensuring consistency between the datasets.
    \item Finally, we construct the merged catalog by treating matched 
    stars as single entries. For these matched stars, already corrected 
    for the zero-point, we calculate a common LOS velocity and its 
    associated uncertainty using the weighted mean of their individual 
    measurements.\footnote{We do not simply adopt the entry from the catalog with the smallest nominal error, as uncertainty-weighted averaging makes optimal use of both measurements and yields the most precise estimate.}
    Stars without matches are retained in the catalog 
    unchanged, aside from the applied zero-point adjustment.
\end{enumerate}

The resulting combined catalog
contains 1,760
Sculptor members with LOS 
velocity uncertainties smaller than the galaxy's intrinsic velocity 
dispersion.\footnote{This threshold is critical for robust mass 
modeling, as highlighted in Paper~I and other studies 
(e.g., \citealt{Vitral+22,Vitral+23}).} This merged dataset represents 
the most comprehensive spectroscopic catalog assembled for any dwarf 
galaxy to date.

\begin{figure}
\centering
\includegraphics[width=\hsize]{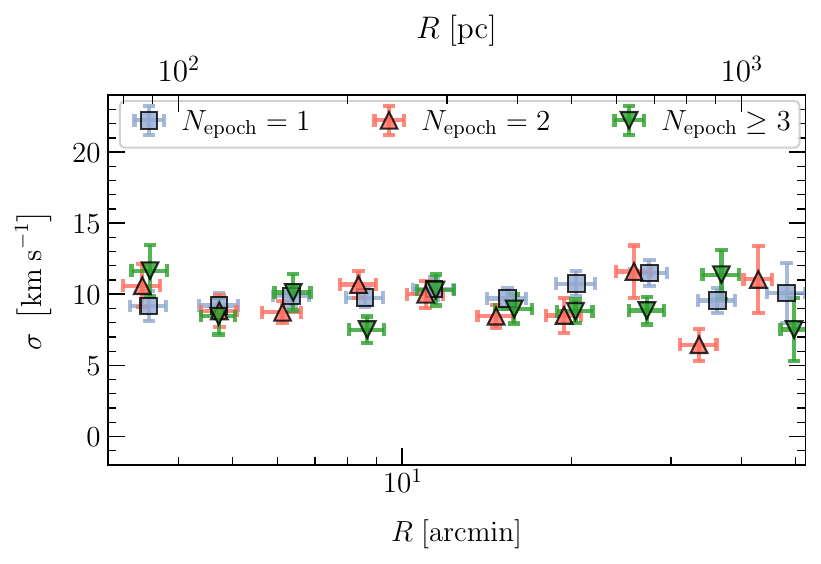}
\caption{\textit{Impact of unresolved binaries:} 
We compare multi-epoch velocity dispersion profiles as a function of projected radii for groups of LOS data categorized by the number of measured epochs. The 
overall agreement within 1-$\sigma$ across different $N_{\rm epoch}$ groups, combined with the fact that the $\sigma_{\rm LOS}$ profiles for higher 
$N_{\rm epoch}$ are not significantly smaller than those for the $N_{\rm epoch} = 1$ subset, provides relative confidence that unresolved binaries do not influence our mass estimates beyond the statistical uncertainties.}
\label{fig: multi-epoch-disp}
\end{figure}

\subsubsection{Binaries} \label{sssec: binaries-los}

In recent years, unresolved binaries in stellar systems have garnered 
increased attention due to their potential role in inflating velocity 
dispersion profiles, which can result in overestimating the mass budget 
\citep[e.g.,][]{Rastello+20,Pianta+22}. However, our previous analyses of 
Draco showed no significant impact from binaries, despite its high binary 
fraction \citep{Spencer+18}. This raises the question of how such inflation patterns might depend on various data characteristics, such as the velocity dispersion 
and the number of measured epochs \citep{Minor+10,Wang+23,Arroyo-Polonio+23}.

Following the approach in Paper~I, we assess the influence of unresolved 
binaries on Sculptor's dynamics by comparing LOS velocity dispersion 
profiles for stars grouped by their number of measured epochs. As 
discussed in our earlier work, if binaries contribute to inflated 
velocity dispersions, we would expect stars with single-epoch LOS 
velocity measurements to exhibit higher dispersions than those with 
multi-epoch measurements, where the effects of binary motion are largely 
averaged out \citep{Wang+23}.

Figure~\ref{fig: multi-epoch-disp} shows that, despite Sculptor's high 
binary fraction,\footnote{\cite{Arroyo-Polonio+23} recently estimated a binary fraction of $0.55^{+0.17}_{-0.19}$ for this dwarf galaxy.} the velocity dispersion profiles derived from stars with different numbers of epochs remain consistent within 1-$\sigma$ uncertainties. This finding aligns with the previous results for Draco. We therefore conclude that Sculptor's binary population does not introduce any bias into our mass-anisotropy estimates beyond the level of statistical uncertainties.

\begin{figure}
\centering
\includegraphics[width=\hsize]{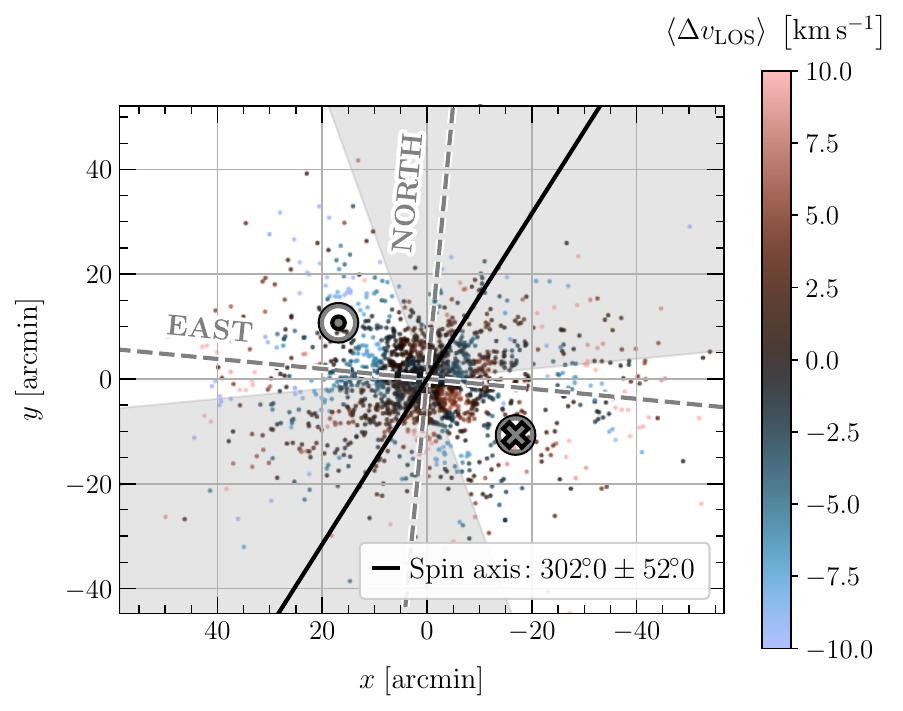}
\caption{\textit{Rotation:} 
Line-of-sight (LOS) rotation fit for Sculptor, with stars represented as 
scattered points, colored by the squared-error weighted mean LOS velocity, within a local circle 
of radius $3\farcm0$. The color bar is offset by the galaxy's bulk LOS 
motion, such that relative receding velocities are shown in red 
(positive, cross symbol) and relative approaching velocities in blue (negative, dot symbol). 
The $x$- and $y$-axes align with the galaxy's photometric major and 
minor axes, respectively, while the East and North directions are 
indicated with dashed gray lines. The fitted rotation axis is 
represented by a solid black line, with the associated 1-$\sigma$ uncertainty from 
bootstrap realizations shown as a gray shaded region.}
\label{fig: rotation}
\end{figure}

\subsubsection{Rotation} \label{sssec: rotation}

Beyond the perspective rotation due to Sculptor's motion in the sky, which we have removed in Section~\ref{ssec: los}, this dSph might still have residual internal rotation that can be observed through the variation of its LOS first-order velocity moments. Indeed, recent studies have suggested that Sculptor carries mild rotation in the LOS \citep[e.g., $\left<v/\sigma\right> = 0.15 \pm 0.15$,][]{Zhu+16}, with a spin axis aligned in between typical oblate/prolate configurations \citep[][figure~6]{Arroyo-Polonio+24}.

While the Jeans modeling described in Section~\ref{ssec: jeans-jampy} 
directly accounts for the LOS first-order velocity moments, we do opt to model Sculptor as an oblate spheroid.\footnote{The Jeans modeling code we use, \jampy, is currently designed to handle either oblate or prolate configurations, but not intermediate arrangements.} To ensure this assumption is 
reasonable, we first analyze Sculptor's rotation profile. This analysis includes determining the direction of the rotation axis and the LOS velocity amplitude along the galaxy's photometric major axis, allowing for further comparison with our Jeans fits.

We model rotation following a similar approach to that in Paper~I, by 
defining eight concentric annuli with linearly-spaced projected radii.
For each 
annulus, we fit a sinusoidal function to the LOS velocity data. While 
each annulus is assigned an independent amplitude, we enforce a common offset and phase across all annuli to recover the bulk LOS velocity and 
the global spin axis, respectively. 
In our formalism, we first project the galaxy onto Cartesian coordinates 
on the sky (as in eq.~2 from \citealt{GaiaCollaborationHelmi+18}), and then align it with Sculptor's major and minor axes, measuring angles clockwise from the positive $x$-axis 
(associated with an angle of $0^\circ$) to the positive $y$-axis. 
\footnote{For Sculptor, these directions approximately align with the 
East and North directions, respectively.} To better estimate the 
uncertainties in our fits, we employ a bootstrap methodology with 
10,000 realizations.

Figure~\ref{fig: rotation} shows the fitted rotation axis direction 
alongside a smoothed map of LOS velocities for the stars included in our 
modeling. We determine a rotation axis direction of $\phi_{0} = 302\fdg0 \pm 52\fdg0$, consistent with an oblate configuration 
(i.e., $\phi_{0} = 270\fdg0$). This result is in 
good agreement with recent estimates by \cite{Arroyo-Polonio+24}, whose 
spin axis direction, expressed within our formalism, corresponds to 
$331\fdg4 \pm 18\fdg0$. The rotation amplitude along the photometric major 
axis is presented later in Figure~\ref{fig: jeans-radial-profiles} together with our PM data, but a brief look reveals 
a modest but non-negligible rotation signal of $\sim 2~\kms$, which amounts to approximately 20\% of Sculptor's internal velocity dispersion. This low rotation amplitude also explains why the rotation axis direction remains poorly constrained in our fits.

\begin{figure}
\centering
\includegraphics[width=0.8\hsize]{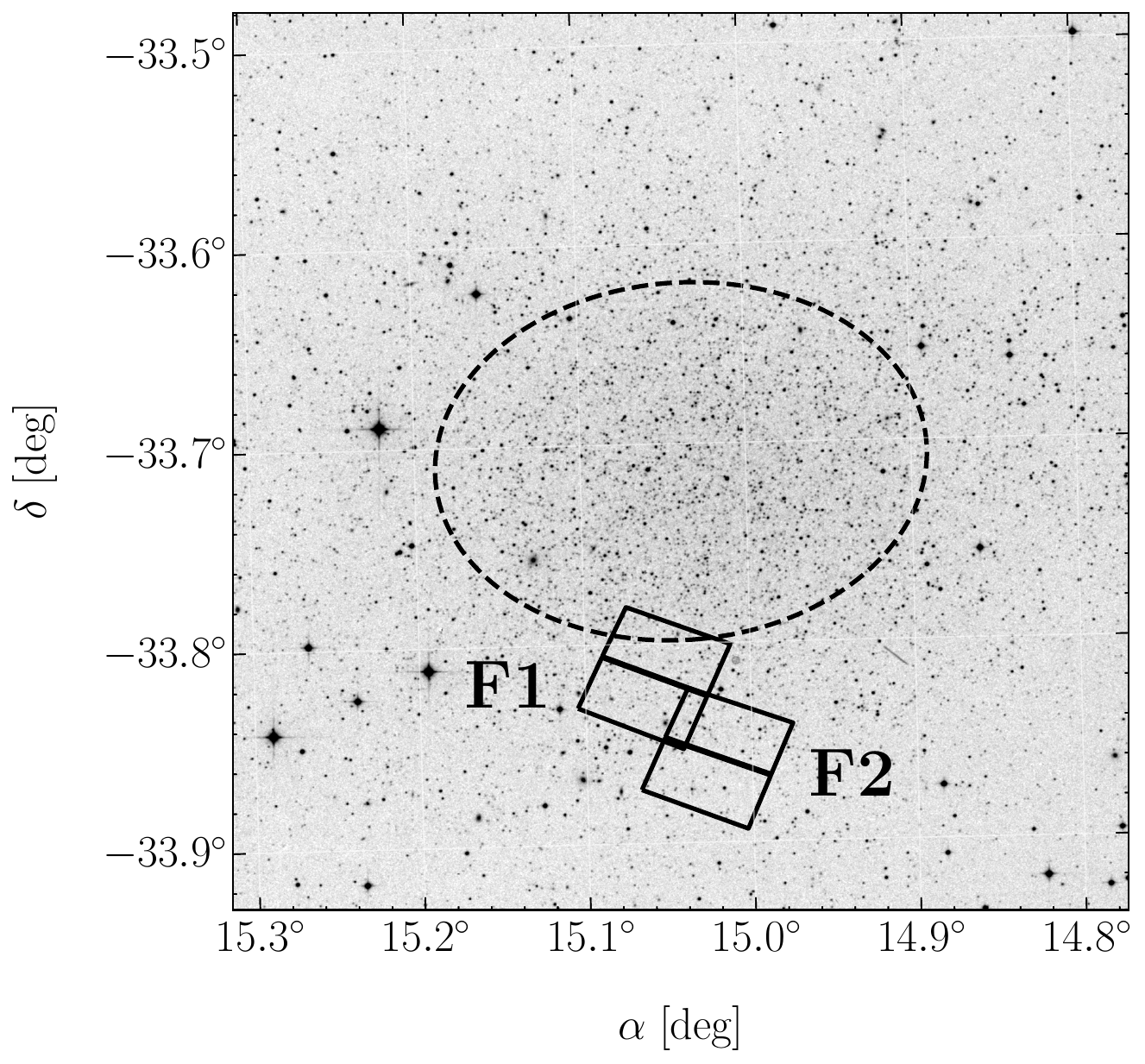}
\caption{\textit{Observed fields:} \hst\ target fields (background image is from the STScI Digitized Sky Survey, see acknowledgments), with a black ellipse showing Sculptor’s half-number radius.}
\label{fig: observed-fields-dss}
\end{figure}

\subsection{Proper Motions} \label{ssec: pms}

\subsubsection{Observations and Astrometric Catalogs} \label{sssec: obs}

For our new PM measurements of Sculptor stars, we utilized multi-epoch 
\hst\ ACS/WFC imaging data. Details about the field locations and earlier 
epoch observations for our target fields, F1 and F2, are provided in 
\citet{Sohn+17}. The locations of these fields are also shown in 
Figure~\ref{fig: observed-fields-dss}. In brief, both fields had two 
epochs of imaging data obtained in 2002 and 2013. Both 2002 and 2013 
observations were performed using the F775W filter, and for
2013 observations, an additional filter F606W was used to create CMDs.
These fields were revisited in October 2022 and October 2023 through our \hst\ 
program GO-16737 \citep{2021hst..prop16737S}, using the same telescope 
pointing and orientation as in the previous epochs. During this latest epoch, 
we acquired 12 individual exposures using the F775W filter for each field, 
with each exposure lasting 495 seconds.  

Data analysis largely followed the procedures outlined in 
\citet{Bellini+2018}, \citet{Libralato+2018}, and Paper~I. Here, we 
provide a high-level summary of the PM derivation process and refer the 
reader to those works for methodological details. We downloaded the 
flat-fielded {\tt \_flt.fits} images for all target fields and epochs 
from the Mikulski Archive for Space Telescopes (MAST). These were 
processed using the {\tt hst1pass} program \citep{Anderson2022} to 
derive positions and fluxes for each star in every exposure. Instead of 
using the {\tt \_flc.fits} images, which are corrected for charge 
transfer efficiency (CTE) losses, we employed the table-based CTE 
correction option in {\tt hst1pass}. This is an improved version of the 
corrections used in earlier works (\citealt{Anderson2022}; Anderson, 
in prep.). Positional corrections were applied using the ACS/WFC 
geometric distortion solutions from \citet{Kozhurina-Platais+2015}, 
further extended to account for time-dependent distortion variations 
beyond 2020 (V. Kozhurina-Platais, private communication).  

For each field, a ``master frame" was constructed using the average 
positions of stars from the repeated first-epoch exposures. The 
($X, Y$) axes of these master frames were aligned to ($\alpha, \delta$) 
coordinates by registering stellar positions to the \gaia\ DR3 
astrometric system. Positions of stars from subsequent epochs were 
aligned to these master frames using a six-parameter linear 
transformation, and average positions were computed for each epoch. 
This procedure inherently aligns the star fields across epochs, 
resulting in zero PM on average for Sculptor dSph stars. Since our focus 
is on measuring internal velocity dispersion in the plane of the sky, 
this alignment does not affect our results. Uncertainties on 
average positions were determined as the root mean square of repeated 
measurements divided by the square root of the number of exposures.  
Finally, for each field and epoch, we prepared a catalog containing 
star positions as derived above, along with average instrumental 
magnitudes\footnote{The instrumental magnitude in a given filter is 
defined as ${\rm mag} = -2.5 \log c$, where $c$ is the electron count 
per exposure for a source.} in F606W and F775W (from previous epochs), 
as output by {\tt hst1pass}.

\begin{figure}
\centering
\includegraphics[width=\hsize]{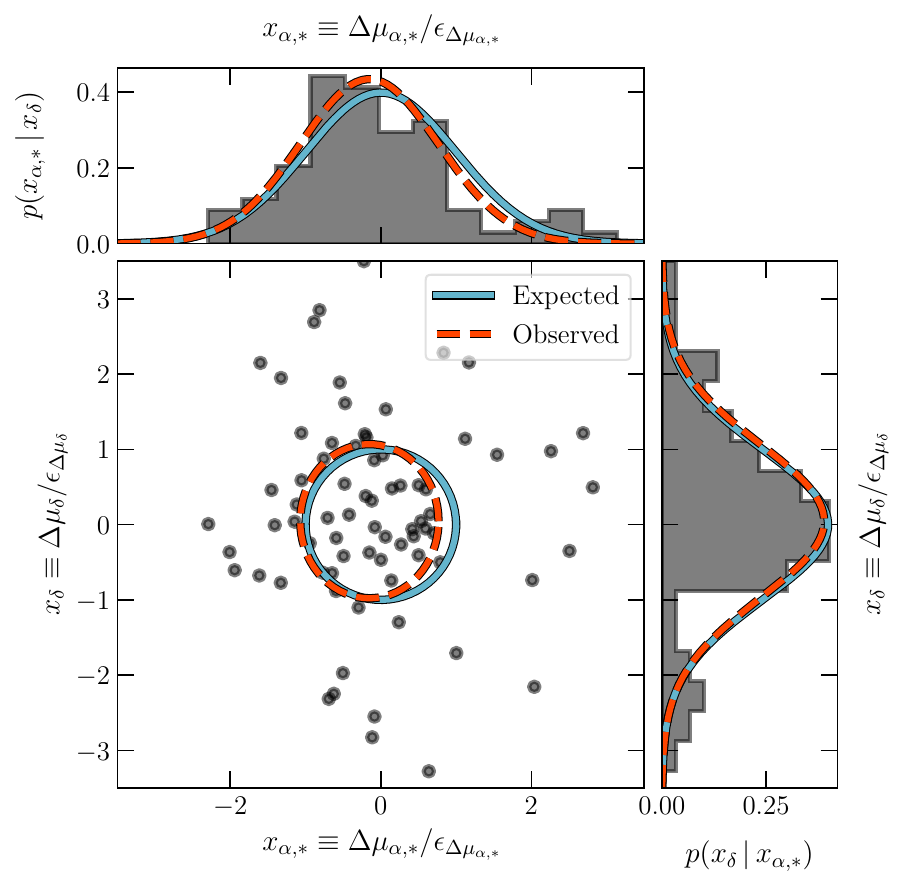}
\caption{\textit{Validation of local corrections:} Histogram of proper motion  
differences for matched stars in overlapping fields, derived from an independent correction process  
for \hst\ systematics and normalized by the combined error budget propagated  
through our procedure. Each axis corresponds to the normalized difference in  
one of the proper motion components, $\mu_{\alpha,*}$ or $\mu_{\delta}$. The dashed-red curve represents a Gaussian constructed using median-based statistics, accounting  
for low-number statistics and the absence of data cleaning. This is compared with  
the expected normal distribution for a well-behaved dataset, shown in blue. The  
agreement between the curves demonstrates the effectiveness of our local corrections,  
as detailed in Section~\ref{sssec: loc-corr-implementation}.}  
\label{fig: intersection-comp}
\end{figure}

\subsubsection{Photometric Cleaning \& Local Corrections} \label{sssec: loc-corr-implementation}

With the catalogs  catalogs of star positions, we compute raw PMs through a least-squares line fit of the master frame ($X, Y$) positions as a function of the epoch time. We use the \textsc{polyfit} routine from \textsc{NumPy} \citep{vanderWalt11}, assuming measured ($X, Y$) uncertainties, and no $\chi^{2}$ re-scaling.
Next, the procedures for cleaning the photometry, removing CMD outliers, and applying local PM corrections to account for \hst\ systematics followed the methodology outlined in Paper~I (sections~2.3.2 and 2.3.3). Below, we briefly summarize the key steps, which are applied independently to each field.

\begin{figure}
\centering
\includegraphics[width=\hsize]{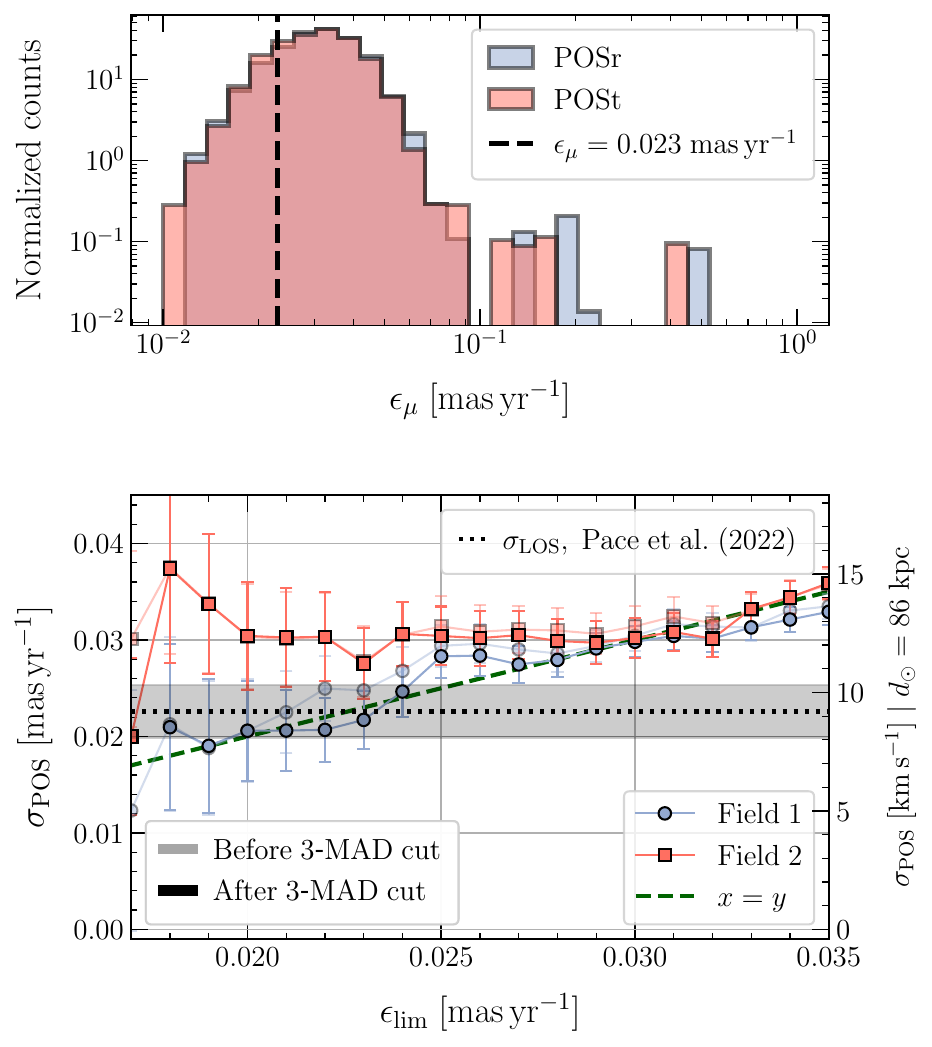}
\caption{\textit{Proper motion errors:}
The upper panel shows the normalized histogram of proper motion uncertainties in the radial (blue) and tangential (red) directions of the plane of the sky. The black dashed vertical line marks the threshold used to clean the data. The motivation for this threshold is illustrated in the lower panel, following a format similar to Figure~9 from Paper~I. There, we plot the measured proper motion dispersion, $\sigma_{\mathrm{POS}}$, as a function of the maximum proper motion uncertainty in the dataset, $\epsilon_{\mathrm{lim}}$. Here, blue curves correspond to Field~1, and red curves to Field~2. Transparent curves show results before applying the 3-MAD cleaning described in Section~\ref{sssec: loc-corr-implementation}, while solid curves reflect the cleaned subsets. For reference, a dashed green line marks the identity relation ($x = y$). The plot demonstrates the importance of imposing a threshold at $\epsilon_{\mathrm{lim}} \sim 0.023~\masyr$: including stars with proper motion uncertainties larger than the galaxy’s intrinsic dispersion leads to a systematic overestimation of $\sigma_{\mathrm{POS}}$. For context, the lower panel also shows the galaxy's line-of-sight velocity dispersion from \citet{Pace+22}, along with the conversion from $\masyr$ to $\kms$, assuming a heliocentric distance of 86~kpc.}
\label{fig: pm-error-analyses}
\end{figure}

\begin{enumerate}
    \item We begin by identifying and removing clear outliers in the CMD and in 
    the magnitude-\texttt{QFIT}\footnote{\texttt{QFIT} quantifies a 
    combined measure of goodness-of-fit and S/N; see 
    \citet{Anderson+06,Libralato+14} for details.} diagrams, employing a 
    friends-of-friends algorithm.

    \item Next, we address \hst\ instrument systematics, which often appear as spatially-correlated mean motions of stars across different regions of the telescope detectors \citep[e.g.][]{Bellini+2018,Libralato+2018}. To correct for this, we calculate and apply median-based shifts, with uncertainties propagated to both PM directions, using a local network of neighboring stars.\footnote{For clarity, this correction is not applied iteratively but only once, during the final stage of the PM construction. As described in Paper~I, the correction introduces an additional spread in the data, which is accounted for in the final PM uncertainty through a quadrature sum.}
    To assess the reliability of this correction, we performed a dedicated Monte Carlo test, where we generated mock stellar fields with realistic uncertainties and injected spatial distortions mimicking detector systematics.\footnote{The injected spatial distortions were constructed using a sum of Fourier modes across a mock stellar field. The procedure was repeated thousands of times with new realizations of velocity fields, star positions, and distortion phases to ensure statistical significance.} We then applied our standard local correction procedure and compared the recovered velocity dispersions to the known input values. 
    The test showed that the propagated uncertainties in our
    locally-corrected estimates are realistic, despite any
    unmodeled correlations that the corrections may introduce.
    Moreover, we find that the locally-corrected dispersions are
    unbiased to with a fraction of these propagated uncertainties.
    Hence, the final results of our work are insensitive to the
    details of our local corrections.

    \item The effectiveness of these corrections is validated through diagnostic 
    checks, such as reproducing the equivalent of figure~8 from Paper~I. 
    This process effectively removes first-order velocity moments, such as rotation 
    or global streaming motions, while preserving the second-order 
    moments necessary for accurate mass-anisotropy modeling. We also test the impact of this data manipulation later on Section~\ref{sssec: mocks-lcorr-lnlik}.
\end{enumerate}

A key feature of our new data is that, while each field  
is initially processed independently, they include overlapping regions,  
as shown in Figure~\ref{fig: observed-fields-dss}. This intersection  
allows for a direct comparison of matched stars. If the local corrections are effective and properly applied, the overall PM differences between matched
stars should be consistent with their associated PM uncertainties and  
should not exhibit significant systematic shifts. Specifically, the  
distribution of PM differences, normalized by their combined error budget,  
is expected to approximate a Gaussian distribution with a mean of zero  
and a standard deviation of one.

To test this, we account for the non-Gaussian nature of the datasets,  
arising primarily from two factors: the absence of error cleaning in this  
initial subset of matched stars and the low-number statistics. These  
factors can influence the distribution of PM differences, potentially  
introducing asymmetries or heavier tails. To address these issues, we  
employ median-based statistics to constrain the center and spread of the  
data. Specifically, we define  
\begin{equation}  
    x_{i} \equiv \Delta \mu_{i} / \epsilon_{\Delta \mu_{i}} \ ,  
\end{equation}  
where the suffix $i$ stands for $\alpha, *$ or $\delta$\footnote{Hereafter,  
$\alpha, *$ refers to the component $\mu_{\alpha, *} = ({\rm d}\alpha / {\rm d}t)  
\cos{\delta}$ from the PM vector.} and $\epsilon$ is used to refer to uncertainties, rather than the $\sigma$ symbol used for the standard deviation. Next, we use the median and the  
median absolute deviation (MAD) of the $x_{i}$ distribution to  
construct a corresponding Gaussian curve. The MAD, corrected to account  
for Gaussian standard deviations as defined in \cite*{Beers+90}, is  
\begin{equation}  
    {\rm MAD} = \frac{\eta_{50}(|x - \eta_{50}(x)|)}{0.6745}  
\end{equation}  
where $\eta_{50}(x)$ refers to the 50th percentile (i.e., the median) of  
the distribution of $x$.

Figure~\ref{fig: intersection-comp} compares the Gaussian curve computed  
using the median and MAD (shown in red) with the expected normal  
distribution (shown in blue). The agreement between these two curves  
supports the validity of the local corrections applied to our PM dataset.  
Not only are the PM differences not significantly shifted toward positive  
or negative values, but the propagated uncertainties also adequately  
encompass the observed spread in the data. 
As a further check, we also performed this analysis using a skewed Gaussian fit,\footnote{This test was motivated by the prominent tail towards ($x_{\alpha,*}$, $x_{\delta}$) $ \sim (2.5, -2.5)$ visible in Figure~\ref{fig: intersection-comp}.} finding that the resulting mean and dispersion remained consistent with those of the expected normal distribution.
Following this validation, we  
proceed to the next stage, which involves cleaning stars with high PM errors and removing outliers in PM.

\subsubsection{Outliers and Underestimated Errors} \label{sssec: under-err}

Before constructing PM profiles, we perform a final  round of data cleaning for our PM measurements, guided by the analyses  presented in Paper~I. Specifically, we compute the global velocity  dispersion\footnote{Here and throughout this work, the dispersion of a  discrete dataset with uneven uncertainties is computed as described in  appendix~A of \cite{vanderMarel&Anderson&Anderson10}, a procedure  thoroughly validated in multiple works (e.g., \citealt{Watkins+15-disp,Vitral+23} and Paper~I).} of our PM dataset for each field, applying  various thresholds for the maximum allowable PM uncertainty. We select  the highest threshold that meets two criteria: (i) it remains within the same  order of magnitude as the expected LOS velocity  dispersion, for a given distance, and (ii) it lies in a plateau of velocity dispersions before increasing due to the inclusion of measurements with large formal errors, which -- if underestimated -- can bias the results. These criteria can be better visualized in Figure~\ref{fig: pm-error-analyses}.
As discussed in detail in Paper~I and prior works (e.g. \citealt{Watkins+15-disp, Vitral+22}), this step is critical for accurate mass modeling. Large formal errors, particularly when underestimated and exceeding the intrinsic velocity dispersion, can introduce significant biases into equilibrium-based mass measurements.

Based on this, we adopt a threshold of $\epsilon_{\mu} <  
0.023~\masyr$ on the PM error;\footnote{This translates to a transverse velocity of $9.4~\kms$, for a given distance of $85.8$~kpc (cf. Table~\ref{tab: mass-modeling-axi}).} stars with larger errors are excluded from the subsequent analysis. Additionally, we apply a 3-$\sigma$ outlier cut, as in  
Paper~I. We once again use median-based statistics (median and MAD) rather than the nominal mean and standard  
deviation to ensure robustness against outliers.\footnote{Note that this step mainly serves to exclude stars too far from the galaxy's bulk motion; in practice, only one of the stars removed lay within 5-$\sigma$ from the median.} Finally, we treat  
matched stars from overlapping fields as single entries, following a  
similar approach to the merging of our LOS datasets (i.e. using the weighted mean of the individual measurements).
Hence, after applying the rigorous selection criteria described throughout this section, we retained a high-quality subset of 119 stars from the original \hst\ observations,\footnote{For context, the original subset contained 8,317 and 5,087 stars in Fields~1 and 2, respectively, while the field-merged PM subset with no error cleaning whatsoever contained 2,070 stars.} all of which having proper-motion uncertainties below the intrinsic velocity dispersion. This yields the most precise PM dataset ever assembled for this dwarf galaxy.

\begin{figure*}
\centering
\includegraphics[width=0.9\hsize]{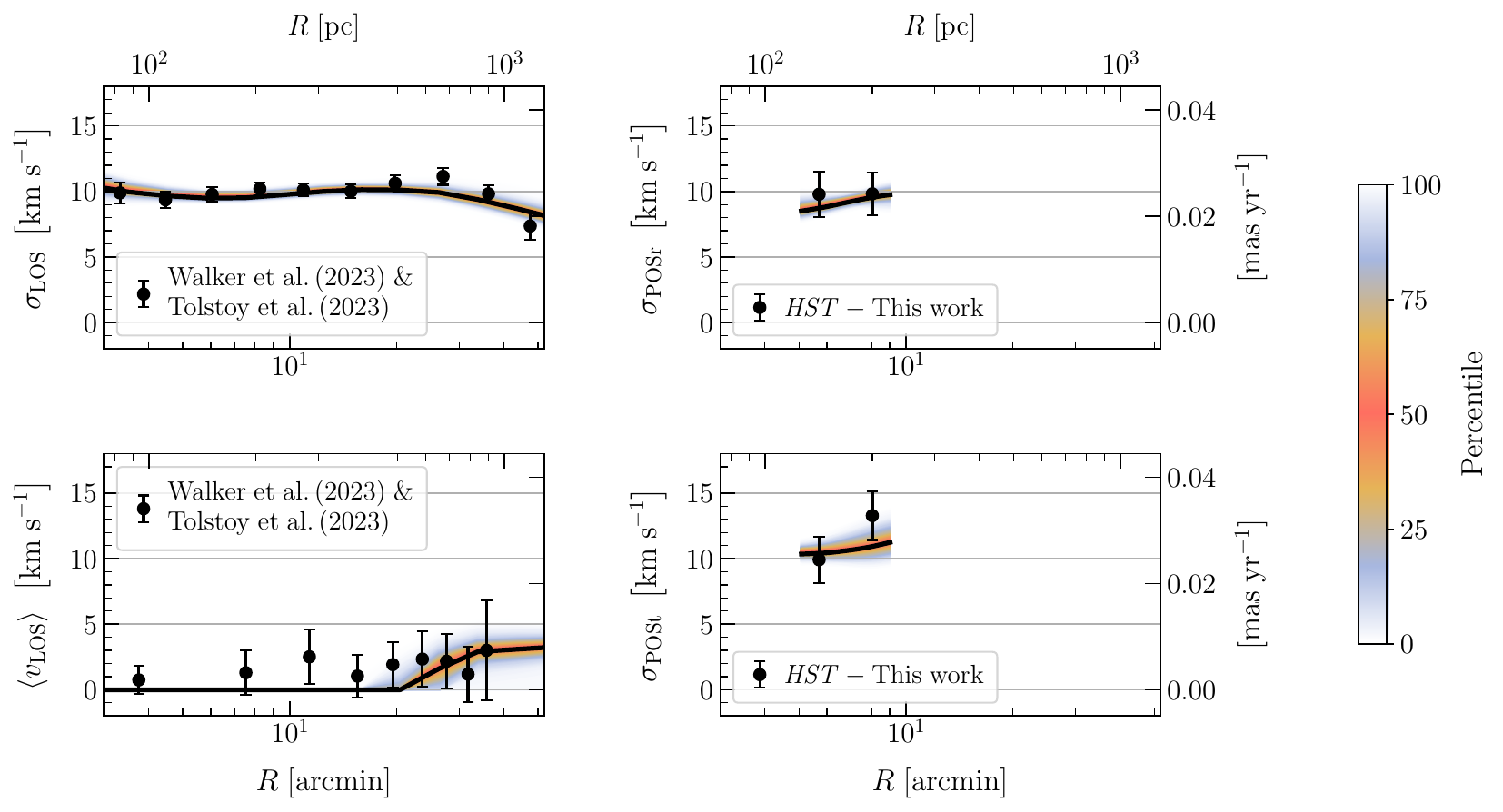}
\caption{\textit{Radially dependent kinematics:} Setup of the kinematic data used in our Jeans modeling, shown as black dots. While binned profiles are displayed for clarity, the modeling uses a likelihood approach on a discrete dataset. The upper and lower left panels show the angularly averaged line-of-sight velocity dispersion and major-axis first-order moment computed with data from \citet{Walker+23} and \citet{Tolstoy+23}. The upper and lower right panels show the radial and tangential transverse velocity dispersions from our \hst\ programs, over the fields in Figure~\ref{fig: observed-fields-dss}, with error bars having comparable contributions from shot noise and measurement uncertainties. Solid black lines and shaded regions mark the median and posterior percentiles of our Jeans fits for a fiducial inclination of $57\fdg1$, also used to convert proper motions to velocities in the rightmost panel. Model curves are interpolated over the projected radius $R$ for visualization, though they also vary with projected angle.}
\label{fig: jeans-radial-profiles}
\end{figure*}

\subsubsection{Proper Motion Dispersion Profiles} \label{sssec: PMsigmaprofiles}

With our new PM dataset, we present for the first time in Figure~\ref{fig: jeans-radial-profiles} the radially-resolved transverse velocity dispersion profiles of the Sculptor dSph.\footnote{For context, the error bars on the proper motion dispersion bins reflect comparable contributions from shot noise and measurement uncertainties.}
Although these profiles are less spatially extended than those shown for Draco in Paper~I, they remain valuable for breaking degeneracies between the galaxy’s mass distribution and orbital structure -- that is, the velocity anisotropy, which describes whether the stellar orbits are preferentially radial or tangential.

Quantitatively, we find a ratio of tangential to radial PM dispersions of $\left< \sigma_{\rm POSt} \right> / \left< \sigma_{\rm POSr} \right> = 1.19 \pm 0.19$ across the spatial extent of the data, where POSr and POSt refer to the plane-of-sky radial and tangential directions, respectively. The ratio of LOS to PM dispersions is $\left< \sigma_{\rm LOS} \right> / \left< \sigma_{\rm POS} \right> = 0.93 \pm 0.08$, with $\sigma_{\rm POS}$ denoting the average dispersion over both transverse directions. The former ratio is independent of the galaxy’s distance, while the latter scales inversely with it.\footnote{The quoted value assumes $D = 85.83$~kpc, as predicted by our Jeans fits in the final row of Table~\ref{tab: mass-modeling-axi}.}

These measurements place important constraints on Sculptor’s internal velocity structure, particularly on its velocity anisotropy. As discussed in Sections~\ref{sec: methods} and \ref{sec: results}, the relative values of the velocity dispersions in different directions -- captured here by ratios such as $\left< \sigma_{\rm POSt} \right> / \left< \sigma_{\rm POSr} \right>$ and $\left< \sigma_{\rm LOS} \right> / \left< \sigma_{\rm POS} \right>$ -- provide strong leverage for disentangling the orbital structure and intrinsic shape of the galaxy.

\section{Methods} \label{sec: methods}

\subsection{Axisymmetric Jeans Modeling: \jampy} \label{ssec: jeans-jampy}

As in Paper~I, we use the mass anisotropy modeling code \jampy~\citep{Cappellari20}, configured such that the velocity ellipsoid aligns with spherical coordinates. This choice follows our assumption of a spherical DM potential, which is expected to dominate the system's dynamics when compared to Sculptor's luminous axisymmetric component. In this configuration, \jampy\ applies the Jeans equations for rotating axisymmetric systems (e.g., \citealt*{Bacon+83}),
\begin{subequations}
\begin{flalign}
    &\frac{\partial \left(\nu \left<v^{2}_{r}\right>\right)}{\nu \, \partial r} + \frac{(1 + \beta_{\rm J}) \, \left<v^{2}_{r}\right> - \left<v^{2}_{\phi}\right>}{r} = - \frac{\partial \Phi}{\partial r} \ , \\
    &\frac{(1 - \beta_{\rm J})}{\nu} \left[\frac{\partial \left(\nu \left<v^{2}_{r}\right>\right)}{\partial \theta} + \frac{\nu \left<v^{2}_{r}\right>}{\tan{\theta}}\right] - \frac{ \left<v^{2}_{\phi}\right>}{\tan{\theta}} = - \frac{\partial \Phi}{\partial \theta} \ ,
\end{flalign}
\end{subequations}
where $\Phi$ is the gravitational potential, and $\left<.\right>$ denotes the distribution function-averaged quantity. The anisotropy parameter $\beta_{\rm J}$, assumed to be independent of $\theta$, is given by
\begin{equation} \label{eq: beta-jampy}
    \beta_{\rm J} \equiv 1 - \frac{\left<v^{2}_{\theta}\right>}{\left<v^{2}_{r}\right>} = 1 - \frac{\sigma^{2}_{\theta}}{\sigma^{2}_{r}} \ ,
\end{equation}
where the second equality assumes $\left<v_{\theta}\right> = \left<v_{r}\right> = 0$. Due to symmetry and continuity, axisymmetric models always satisfy $\left< v_{\phi} \right> = 0$ and $\left< v_{\phi}^2 \right> = \left< v_{\theta}^2 \right>$ along the symmetry axis. Consequently, along this axis, the anisotropy parameter $\beta_{\rm B}$, as defined by \cite{Binney80},
\begin{equation}    \label{eq: binney-beta}
    \beta_{\rm B} \equiv 1 - \displaystyle{\frac{\left< v_{\theta}^{2} \right> + \left< v_{\phi}^{2} \right> }{2 \,\left< v_{r}^{2} \right> }} \ ,
\end{equation}
equals $\beta_{\rm J}$. Models with $\beta_{\rm J} = 0$ yield the same predicted second velocity moments as models in which the distribution function $f(E,Lz)$ does not depend on a third integral. Such models have been widely used for fitting data of axisymmetric systems \citep[e.g.][]{vanderMarel91}. Away from the symmetry axis, models with $\beta_{\rm J} = 0$ do {\it not} have an isotropic velocity dispersion tensor. 

By not enforcing $\left<v_{\phi}\right> = 0$ throughout the entire system, \jampy\ allows users to model rotation in the fitted galaxy -- a feature often neglected in previous kinematic models of Sculptor \citep[e.g.,][]{Massari+17,Hayashi+20}. The first moment in the $\phi$ direction is related to the second-order moment in the radial direction through
\begin{subequations} \label{eq: rotation-jampy}
\begin{flalign}
    \left<v_{\phi}\right>^{2} &= \left<v^{2}_{\phi}\right> - \sigma^{2}_{\phi} \ , \\
    \sigma^{2}_{\phi} &= (1 - \Omega) \, \left<v_{r}^{2}\right> \ ,
\end{flalign}
\end{subequations}
where $\Omega$ is the rotation parameter, also assumed to be independent of $\theta$. This parameter was denoted as $\gamma$ in \cite{Cappellari20}, but we adopt a different notation here to avoid confusion with the inner slope of the DM mass density, which is commonly represented by the same symbol.

From the solutions to these equations, \jampy\ calculates projected velocity moments, which we then use to compute the respective quantities in the LOS and POS directions. We sample the parameter space using the Markov Chain Monte Carlo (MCMC) sampler \textsc{emcee} \citep{emcee}, by using a likelihood $\mathcal{L}$ defined as
\begin{flalign}
     \label{eq: log-lik}
    {\cal L} &= \prod_{i, \rm LOS} G_{\rm LOS}(v_{{\rm LOS}, i})  \nonumber \\
    & \times \prod_{i, \rm POS} G_{\rm POSr}(\mu_{{\rm POSr}, i}) \, G_{\rm POSt}(\mu_{{\rm POSt}, i}) \ .
\end{flalign}

With respect to equation~\ref{eq: log-lik}, a few key considerations must be clarified:
\begin{itemize}
    \item $G(x)$ represents a Gaussian function of argument $x$.
    \item Since the stars with PMs in our dataset do not have a corresponding $v_{\rm LOS}$ measurement,\footnote{This is because our \hst\ exposures target the more numerous faint stars, while the brighter stars for which ground-based spectroscopic measurements are typically possible are often saturated in the respective \hst\ images.} the product of the POS Gaussian components can be effectively treated separately from the LOS Gaussian component, with no risk of neglecting cross-terms in spherical or axisymmetric symmetry.
    \item For all Gaussian components, the respective standard deviation is derived from the output of \jampy\ along the corresponding dimension, using the relation $\sigma^{2} = \left< v^{2} \right> - \left< v \right>^{2}$. To this, we add the individual stellar errors quadratically, assuming Gaussian-distributed uncertainties.
    \item For the LOS component, the Gaussian mean is taken as the bulk LOS motion of Sculptor, plus the respective first-order moment estimated by \jampy, and adjusted to the pre-defined sense of rotation in Sculptor. For the POS components, however, we adopt a null mean as a result of our local cleaning process explained in Section~\ref{sssec: loc-corr-implementation}, which effectively removes streaming motions in our PMs.
    \item Although the assumption of Gaussian-distributed projected velocities is formally incorrect in the presence of anisotropy \citep[][figure~8]{Merritt87}, we retain it due to the demonstrated convergence of this approach when applied to mock datasets generated by self-consistent distribution functions \citep{Read+21}, as well as its consistency with results from well-studied dynamical systems \citep[see][for an application to $\omega$ Centauri]{Watkins+13}.
\end{itemize}

Throughout our posterior sampling, we employ priors as defined in Section~\ref{sssec: priors-jampy}. Our MCMC routine is configured to run a minimum of 20\,000 steps per fit,\footnote{If proper convergence is not achieved, we add an additional 10\,000 steps, which was sufficient in our case.} and we execute the runs in parallel on the local cluster at the University of Edinburgh. In these configurations, each run takes approximately 3 days to complete. We repeat the process for a different set of possible inclinations, drawn from the PDF derived in Section~\ref{ssec: density}. To discard the burn-in phase, we remove the first 75\% of the steps from each chain and visually check that the chains remain stable thereafter.

\subsubsection{Parametrizations \& Priors of \jampy} \label{sssec: priors-jampy}

Our parametrization follows the approach of Paper~I. The baryonic content is fixed and modeled with a flattened Plummer profile, with parameters determined in Section~\ref{ssec: density}. Owing to the lack of external constraints on the halo shape, the DM halo is modeled as a spherical generalized Plummer profile,\footnote{A special case of the $\alpha\beta\gamma$ profile \citep{Zhao96}, with $\alpha = 2$, $\beta = 5$, and a free inner slope $\gamma$ (see eq.~A3 from Paper~I).} with a steeper outer slope ($\beta = 5$) motivated by tidal interactions inferred from Sculptor’s orbital history \citep{Sohn+17}, consistent with the findings of \cite{Penarrubia+10}. This DM parametrization was also adopted in Paper~I, \cite{Hayashi+20}, and \cite{Goldstein&Strigari25}.

This density model depends on three parameters: a scale radius $r_{\rm dark}$, a total DM mass ($M_{\rm dark}$), and an inner slope $\gamma_{\rm dark}$. We assume flat/log-flat priors for all of these variables,
\begin{itemize}
    \item $\gamma_{\rm dark} \in [-2, 2]$, which includes both cuspy ($\gamma = -1$) and cored ($\gamma = 0$) profiles. We allow for positive slopes to account for potential physical mechanisms not necessarily anticipated by $\Lambda$CDM.
    \item $\log{(M_{\rm dark} / [\msun])} \in [6, 12]$. \cite{Read+17} extrapolated classical $M_{\star} - M_{200}$ relations to lower-mass dSphs, such that Sculptor, with a luminous mass of $\sim 8.08 \times 10^{6}~\msun$, should have $M_{200} \sim 10^{9}~\msun$. Therefore, our priors largely cover this range.
    \item $\log{(r_{\rm dark} / [{\rm kpc}])} \in [-1, 2]$. Given the expected $M_{200}$ mass from \cite{Read+17},\footnote{We thank Justin Read for sharing his algorithm to compute the precise $M_{200}$ value expected for Sculptor.} the concentration relation from \cite{Dutton&Maccio14} yields a \cite{Navarro+97} profile scale radius of nearly $2$~kpc. Our prior thus covers this with more than an order of magnitude range. 
\end{itemize}

We assume that the anisotropy parameters, $\beta_{\rm J}$ and $\Omega$, are constant and fit their symmetrized quantity,\footnote{$x_{\rm sym}$ ranges monotonically from $-2$ to $+2$, given that $x$ spans similarly from $-\infty$ to $+1$.} defined as $x_{\rm sym} = x / (1 - x/2)$, using a flat prior between $-1.99$ and $1.99$. To assess the impact of this assumption, we have separately tested variations with a radial dependence, such as generalized parametrizations \citep{Osipkov79,Merritt85}, and found no significant effect on our results and conclusions presented further on.

Finally, we apply Gaussian priors for the bulk $v_{\rm LOS}$ of Sculptor, setting $v_{\rm LOS} = 111.2 \pm 0.3~\kms$ \citep{Arroyo-Polonio+24}, and for the distance modulus, defined as $\mu_{0} = 5 \, \log{(D/[{\rm kpc}])} + 10$. We adopt $\mu_{0} = 19.67 \pm 0.05$, which is the weighted mean of measurements reported in \cite[][RGB and TRGB]{Rizzi02}, \cite[][RR-Lyrae on V and I bands]{Pietrzynski+08}, and \cite[][TRGB on J and K bands]{Gorski+11}.

\subsection{Practical Quantities} \label{ssec: obs-param}

As in Paper~I, we compute derived quantities to better interpret the results of our fits, as some quantities can be difficult to assess in their original form. For instance, due to the mass-anisotropy degeneracy, constraining the asymptotic DM density slope is particularly challenging in the absence of 3D velocities near the galaxy center. Instead, a more meaningful approach is to evaluate the DM density slope within the radial range where both PMs and LOS velocities are available. We define this as an effective DM slope:
\begin{equation} \label{eq: gamma-practicle} 
    \Gamma_{\rm dark} \equiv \displaystyle{\frac{\int_{r_{\rm min}}^{r_{\rm max}} \frac{{\rm d} \log{\rho}}{{\rm d} \log{r}} \, \rho(r) \, {\rm d}r}{\int_{r_{\rm min}}^{r_{\rm max}} \rho(r) \, {\rm d}r}} \ , 
\end{equation}
where $\rho(r)$ is the DM density. The lower bound, $r_{\rm min}$, corresponds to the smallest projected radius in the data where PM information is available. We set the upper bound as $r_{\rm max} = {\rm min}(R_{\rm max, PM}, r_{\Lambda{\rm CDM}}/3)$, where $R_{\rm max, PM}$ is the largest projected radius with available PM data, and $r_{\Lambda{\rm CDM}}$ is the NFW scale radius expected for DM halos in $\Lambda$CDM simulations.\footnote{We adopt this reference value to facilitate comparison with theoretical predictions. The factor $(1/3)$ is introduced to exclude the region where the density profile steepens due to the transition from the inner to outer profile.} (computed with the relations from \citealt{Read+17}).

Additionally, as in our previous work, we compute the circular velocity:
\begin{equation} \label{eq: vcirc} 
    v_{\rm circ}(r) = \sqrt{\frac{G M(r)}{r}} \ , 
\end{equation}
which depends on the total\footnote{The total mass includes both luminous and dark components.} cumulative mass enclosed within radius $r$ and the gravitational constant $G$.

\begin{deluxetable*}{rrlrllrrlr}
\tablecaption{Main results of the \jampy\ axisymmetric Jeans modeling.}
\label{tab: mass-modeling-axi}
\tablewidth{750pt}
\renewcommand{\arraystretch}{1.3}
\tabcolsep=4pt
\tabletypesize{\scriptsize}
\tablehead{
\colhead{$i$} & 
\colhead{$D$} &
\colhead{$\Omega$} &
\colhead{$\beta_{\rm J}$} &
\colhead{$r_{\rm dark}$} &
\colhead{$M_{\rm dark}^{R_{\rm max}}$} &
\colhead{$\gamma_{\rm dark}$} &
\colhead{$\Gamma_{\rm dark}$} &
\colhead{$\overline{\beta_{\rm B}}$} &
\colhead{$v_{\rm circ}^{R_{\rm max}}$} \\
\colhead{[$^{\circ}$]} &
\colhead{[kpc]} &
\colhead{} &
\colhead{} &
\colhead{$\left[10^{2}~\rm{pc}\right]$} &
\colhead{$\left[10^{8}~\msun\right]$} &
\colhead{} &
\colhead{} &
\colhead{} &
\colhead{$\left[\rm{km \, s^{-1}}\right]$} \\
\colhead{(1)} &
\colhead{(2)} & 
\colhead{(3)} & 
\colhead{(4)} & 
\colhead{(5)} & 
\colhead{(6)} & 
\colhead{(7)} & 
\colhead{(8)} &
\colhead{(9)} &
\colhead{(10)} \\
}
\startdata
$  43.7 $ & $  87.69^{+ 1.92}_{- 1.96} $ & $ -2.20^{+ 0.32}_{- 0.37} $ & $  0.98^{+ 0.00}_{- 0.00} $ & $  3.32^{+ 0.55}_{- 0.32} $ & $  1.68^{+ 0.16}_{- 0.14} $ & $  1.35^{+ 0.45}_{- 0.62} $ & $ -0.16^{+ 0.15}_{- 0.20} $ & $ ~~ \, 0.94^{+ 0.00}_{- 0.00} $ & $  20.79^{+ 0.87}_{- 0.79} $ \\
$  47.7 $ & $  85.98^{+ 1.94}_{- 1.94} $ & $ -0.53^{+ 0.19}_{- 0.20} $ & $  0.92^{+ 0.01}_{- 0.01} $ & $  4.26^{+ 0.42}_{- 0.32} $ & $  2.11^{+ 0.25}_{- 0.21} $ & $  1.64^{+ 0.23}_{- 0.45} $ & $  0.52^{+ 0.19}_{- 0.22} $ & $ ~~ \, 0.82^{+ 0.02}_{- 0.03} $ & $  23.49^{+ 1.13}_{- 1.26} $ \\
$  52.4 $ & $  85.79^{+ 2.40}_{- 1.88} $ & $ -0.42^{+ 0.26}_{- 0.24} $ & $  0.86^{+ 0.03}_{- 0.03} $ & $  4.49^{+ 0.69}_{- 0.42} $ & $  2.04^{+ 0.24}_{- 0.27} $ & $  1.58^{+ 0.28}_{- 0.37} $ & $  0.57^{+ 0.19}_{- 0.20} $ & $ ~~ \, 0.68^{+ 0.06}_{- 0.07} $ & $  22.80^{+ 1.49}_{- 1.26} $ \\
$  57.1 $ & $  85.46^{+ 2.07}_{- 1.81} $ & $ -0.91^{+ 0.45}_{- 0.75} $ & $  0.72^{+ 0.07}_{- 0.12} $ & $  4.06^{+ 0.65}_{- 0.51} $ & $  1.50^{+ 0.32}_{- 0.30} $ & $  1.45^{+ 0.41}_{- 0.68} $ & $  0.29^{+ 0.31}_{- 0.41} $ & $ ~~ \, 0.35^{+ 0.17}_{- 0.39} $ & $  19.88^{+ 1.97}_{- 1.93} $ \\
$  61.8 $ & $  85.78^{+ 1.79}_{- 1.92} $ & $ -4.76^{+ 3.08}_{- 4.75} $ & $ -0.10^{+ 0.60}_{- 1.06} $ & $  6.37^{+ 10.14}_{- 2.64} $ & $  0.95^{+ 0.25}_{- 0.17} $ & $ -1.12^{+ 2.09}_{- 0.75} $ & $ -1.39^{+ 1.18}_{- 0.51} $ & $ -1.04^{+ 0.99}_{- 2.15} $ & $  16.05^{+ 1.89}_{- 1.42} $ \\
$  66.5 $ & $  85.73^{+ 1.75}_{- 1.73} $ & $ -6.35^{+ 2.64}_{- 2.89} $ & $ -0.79^{+ 0.71}_{- 0.67} $ & $  12.17^{+ 60.81}_{- 5.82} $ & $  0.90^{+ 0.23}_{- 0.18} $ & $ -1.78^{+ 0.65}_{- 0.15} $ & $ -1.83^{+ 0.44}_{- 0.13} $ & $ -1.91^{+ 1.36}_{- 1.39} $ & $  15.67^{+ 1.77}_{- 1.52} $ \\
$  71.2 $ & $  85.69^{+ 2.02}_{- 2.01} $ & $ -5.42^{+ 2.83}_{- 2.65} $ & $ -0.60^{+ 0.72}_{- 0.65} $ & $  9.64^{+ 27.90}_{- 4.84} $ & $  0.88^{+ 0.18}_{- 0.16} $ & $ -1.65^{+ 1.10}_{- 0.28} $ & $ -1.74^{+ 0.69}_{- 0.21} $ & $ -1.92^{+ 1.34}_{- 1.10} $ & $  15.47^{+ 1.46}_{- 1.37} $ \\
$  75.9 $ & $  85.28^{+ 1.87}_{- 1.86} $ & $ -5.48^{+ 1.91}_{- 2.40} $ & $ -0.77^{+ 0.55}_{- 0.57} $ & $  13.05^{+ 51.39}_{- 6.06} $ & $  0.88^{+ 0.18}_{- 0.18} $ & $ -1.79^{+ 0.50}_{- 0.16} $ & $ -1.85^{+ 0.34}_{- 0.12} $ & $ -2.22^{+ 1.04}_{- 0.93} $ & $  15.56^{+ 1.35}_{- 1.57} $ \\
$  80.6 $ & $  85.48^{+ 1.82}_{- 1.83} $ & $ -5.71^{+ 1.83}_{- 5.48} $ & $ -0.83^{+ 0.54}_{- 0.50} $ & $  15.02^{+ 104.45}_{- 7.24} $ & $  0.88^{+ 0.18}_{- 0.18} $ & $ -1.84^{+ 0.42}_{- 0.11} $ & $ -1.87^{+ 0.29}_{- 0.09} $ & $ -2.18^{+ 1.00}_{- 0.88} $ & $  15.48^{+ 1.45}_{- 1.53} $ \\
$  85.3 $ & $  85.45^{+ 1.84}_{- 1.76} $ & $ -5.83^{+ 1.81}_{- 11.00} $ & $ -0.88^{+ 0.48}_{- 0.47} $ & $  16.18^{+ 93.35}_{- 7.63} $ & $  0.89^{+ 0.16}_{- 0.19} $ & $ -1.87^{+ 0.33}_{- 0.09} $ & $ -1.90^{+ 0.24}_{- 0.08} $ & $ -2.25^{+ 1.14}_{- 0.91} $ & $  15.62^{+ 1.21}_{- 1.62} $ \\
$  90.0 $ & $  85.53^{+ 1.90}_{- 1.81} $ & $ -5.47^{+ 1.79}_{- 4.00} $ & $ -0.86^{+ 0.50}_{- 0.51} $ & $  16.64^{+ 173.46}_{- 8.33} $ & $  0.90^{+ 0.18}_{- 0.19} $ & $ -1.86^{+ 0.36}_{- 0.10} $ & $ -1.88^{+ 0.25}_{- 0.09} $ & $ -2.16^{+ 1.23}_{- 0.98} $ & $  15.65^{+ 1.40}_{- 1.66} $ \\
$ \left< . \right>_{i} $ & $  85.83^{+ 2.02}_{- 1.91} $ & $ -3.16^{+ 1.75}_{- 4.96} $ & $  0.13^{+ 0.78}_{- 1.15} $ & $  7.86^{+ 8.51}_{- 7.86} $ & $  1.39^{+ 0.68}_{- 0.58} $ & $ -0.02^{+ 1.76}_{- 1.82} $ & $ -0.67^{+ 1.23}_{- 1.24} $ & $ -0.56^{+ 1.33}_{- 2.00} $ & $  18.88^{+ 4.23}_{- 3.94} $ \\
\enddata
\begin{tablenotes}
\scriptsize
\item \textsc{Notes} --  
Columns are 
\textbf{(1)} Inclination, in degrees ($90\degree$ is edge-on); 
\textbf{(2)} Heliocentric distance, in kpc;
\textbf{(3)} Rotation parameter (see Eq.~[\ref{eq: rotation-jampy}]); 
\textbf{(4)} $\beta_{\rm J}$ velocity anisotropy parameter, as defined in Eq.~(\ref{eq: beta-jampy}); 
\textbf{(5)} Dark matter scale radius, in $10^{2}$~pc;
\textbf{(6)} Dark matter mass at maximum projected data radius, in $10^{8}$~M$_{\odot}$;
\textbf{(7)} Dark matter asymptotic density slope;
\textbf{(8)} Dark matter density slope averaged over the spatial range where PMs are available;
\textbf{(9)} Globally-averaged $\beta_{\rm B}$ velocity anisotropy, as defined in Eq.~(\ref{eq: binney-beta});
\textbf{(10)} Circular velocity at maximum projected data radius, in $\kms$;
The uncertainties are based on the 16th and 84th percentiles of the marginal distributions. 
The last row displays the integrated estimate for each parameter, averaged over the inclination probability distribution of Sculptor, as described in Section~\ref{ssec: axi-inc}.
\end{tablenotes}
\end{deluxetable*}

\section{Results} \label{sec: results}

We ran \jampy\ over a set of eleven inclinations, linearly spaced from $43\degree.7$ to $90\degree$ (edge-on), covering the range of Sculptor's inclination PDF computed in Section~\ref{ssec: density}, to mimic the analyses from Paper~I. The results are presented in Table~\ref{tab: mass-modeling-axi}.
In addition to the anisotropy parameter $\beta_{\rm J}$, defined in Eq.(\ref{eq: beta-jampy}), we also report the globally-averaged Binney anisotropy, $\overline{\beta_{\rm B}}$. The latter is computed by first determining the stellar-mass-weighted second velocity moments, averaged over the entire system, and then substituting them into Eq.(\ref{eq: binney-beta}).

\begin{figure}
\centering
\includegraphics[width=\hsize]{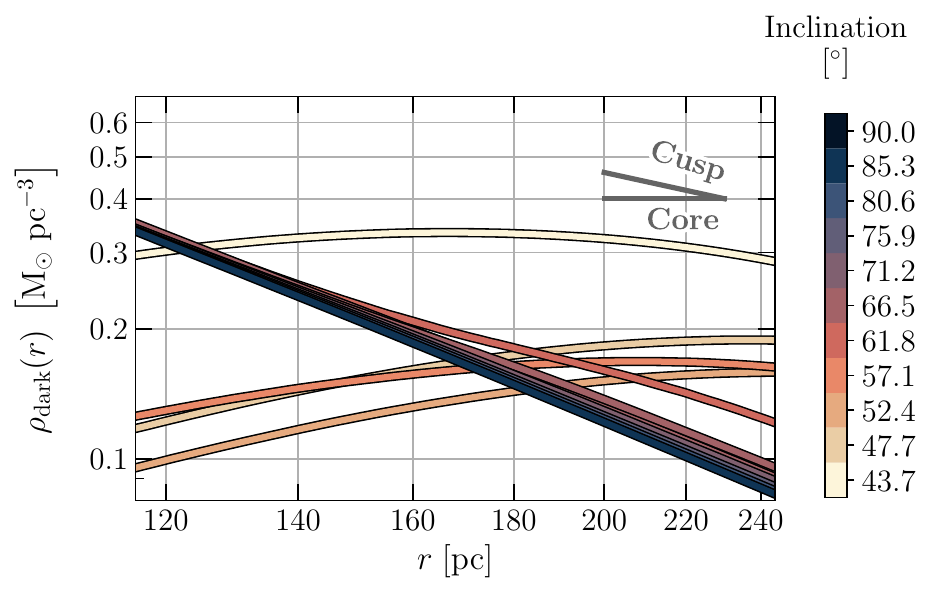}
\caption{\textit{Degeneracy with inclination:} 
Best-fitting dark matter density profiles within the radial range constrained by the three-dimensional velocity data, shown alongside reference slopes representative of core and cusp profiles. Each color corresponds to a tested inclination drawn from Sculptor's inclination probability distribution function. The figure illustrates a pronounced degeneracy between the dark matter density slope and the assumed inclination -- an effect that is overlooked under the assumption of spherical symmetry.} 
\label{fig: rho_vs_inc}
\end{figure}

\subsection{Inclination Dependence} \label{ssec: axi-inc}

Regarding the dependence of model parameters on the adopted galaxy inclination, Paper~I found that, while correlations were present, Draco’s overall parameters remained relatively consistent. That is, the parameters largely retained their signs, preserving the general regime they indicated (e.g., velocity anisotropy remained tangential, and the DM slope differed from the cored case across all inclinations).

In this study, we observe similar correlations with inclination; however, their impact on our modeling conclusions is much more pronounced.
Specifically, the best-fitting flattened systems exhibit extreme radial anisotropy with a corresponding steep density drop, whereas the best-fitting edge-on systems exhibit extreme tangential anisotropy with a corresponding highly-cuspy density profile. This diversity of DM slopes is showcased in Figure~\ref{fig: rho_vs_inc}.

The inclination-averaged best-fit parameters of our model were computed using the inclination PDF shown in Figure~\ref{fig: inc-dist}, following the formalism presented in Paper~I, Section~4.2.1. As expected from the strong degeneracy discussed above, the DM slope and velocity anisotropy are well constrained only for specific inclination values, but not in the averaged case.
This result highlights the critical role of properly accounting for inclination in the mass modeling of flattened systems.

\subsubsection{Higher-order velocity moments} \label{sssec: high-order}

\begin{figure}
\centering
\includegraphics[width=\hsize]{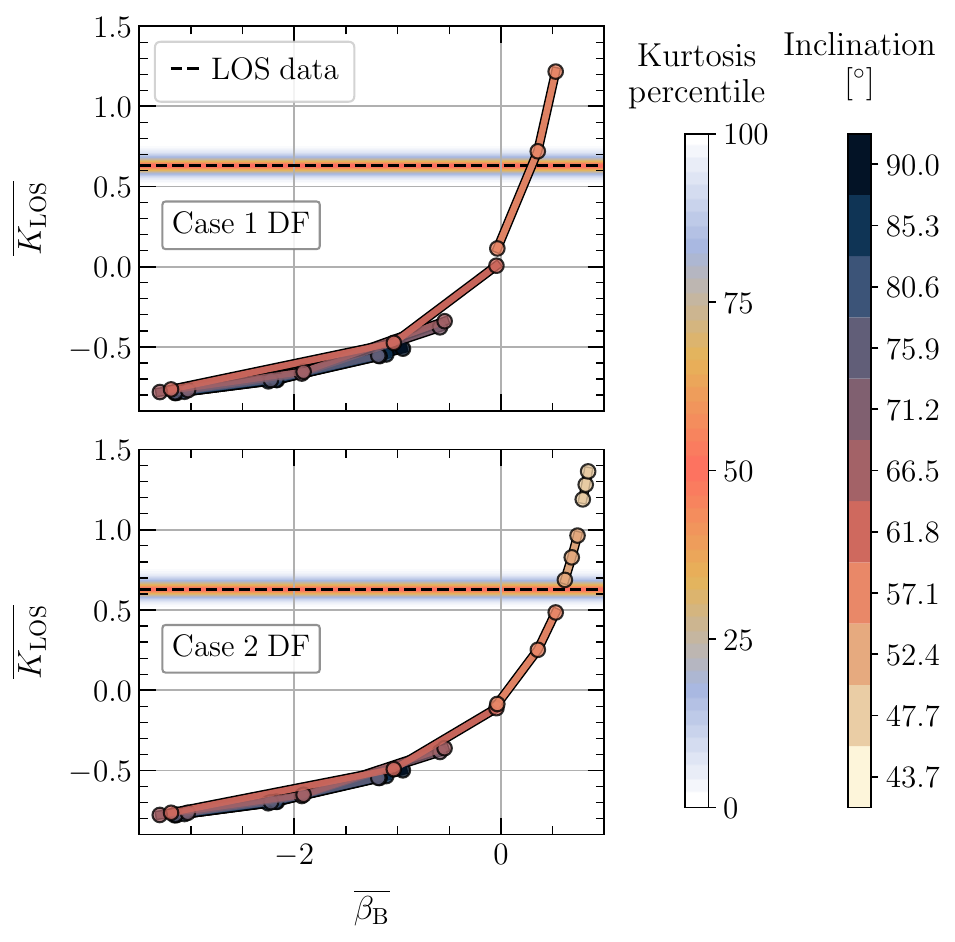}
\caption{\textit{Higher-order moments:} 
Spatially-averaged kurtosis of the line-of-sight velocity distribution as a function of the averaged velocity anisotropy $\overline{\beta_{\rm B}}$, defined in eq.~\ref{eq: binney-beta}. The two panels correspond to different distribution functions explored by the \scf\ software from \cite{deBruijne+96}. The observed kurtosis distribution from our LOS velocity data is shown using the color scale on the left, with the dashed black line marking its mean. For each distribution function and inclination (color scale on the right), we compute the corresponding \scf-predicted kurtosis using three representative values of $\overline{\beta_{\rm B}}$: the 16th percentile, the median, and 84th percentile (from Table~\ref{tab: mass-modeling-axi}, column~9), shown as connected points. The figure illustrates a consistent preference for less inclined (i.e., more flattened) configurations of the Sculptor dSph, regardless of the adopted distribution function.}  
\label{fig: scalefree-kurtosis}
\end{figure}

When first-order velocity moments are available in all three dimensions, the inclination of an axisymmetric system can be directly constrained using relations like those in \citeauthor{Evans&deZeeuw94}~(\citeyear{Evans&deZeeuw94}, eq.~[A4]) or via Jeans modeling with the inclination as a free parameter. However, correcting systematic effects in \hst-based proper motions (as detailed in Section~\ref{sssec: loc-corr-implementation}) removes transverse first-order moments by design. Our Jeans analysis thus relies solely on the LOS first-order moment, which is insufficient to determine inclination on its own.

To address this degeneracy, we turn to the \scf\ models of \citet{deBruijne+96}, further detailed in Paper~I (appendix~B). These models assume an axisymmetric baryonic component in a spherical potential\footnote{This ensures consistency with our current Jeans setup.} and, by adopting fiducial distribution functions and potential forms, offer a reasonable and generalized approximation of systems like dSphs with flattened dispersion profiles. While these models do not capture the detailed radial variation of the logarithmic density and potential slope, they can, unlike Jeans modeling, use higher-order moments of a single velocity dimension (here, LOS) to break degeneracies between mass, anisotropy, and crucially, inclination. This offers educated guesses of the inclination via features such as the kurtosis of the LOS velocity distribution (e.g., Paper~I, fig.~18).

Following this approach, we compute the observed LOS kurtosis to be $\overline{K_{\rm LOS}} = 0.63 \pm 0.05$ and compare it to \scf\ predictions across a range of inclinations, using our inferred velocity anisotropies. Figure~\ref{fig: scalefree-kurtosis} shows two panels, one per \scf\ distribution function type, with the observed kurtosis distribution color-coded as in the leftmost color bar, and its mean marked by a dashed black line. For each inclination, we compute the expected kurtosis from \scf\ using three representative values of the average velocity anisotropy $\overline{\beta_{\rm B}}$: the 16th percentile, the median, and the 84th percentile (from Table~\ref{tab: mass-modeling-axi}, column~9).

In both distribution function cases, the models favor lower inclinations (i.e., more flattened systems). The first panel yields the best match\footnote{Defined by the most evident overlap between the three anisotropy curves and the dashed black line.} at $i = 57\fdg1$, with the second at $i = 52\fdg4$. Although these generalized \scf-based models are not designed for precise inclination fitting, the consistent need for radial velocity anisotropy to match the observed kurtosis supports the idea that Sculptor lies toward the more flattened end of the viable axial ratio range. This result remains robust even under conservative 3- and 4-$\sigma$ clipping of the LOS velocity distribution.

\begin{figure}
\centering
\includegraphics[width=\hsize]{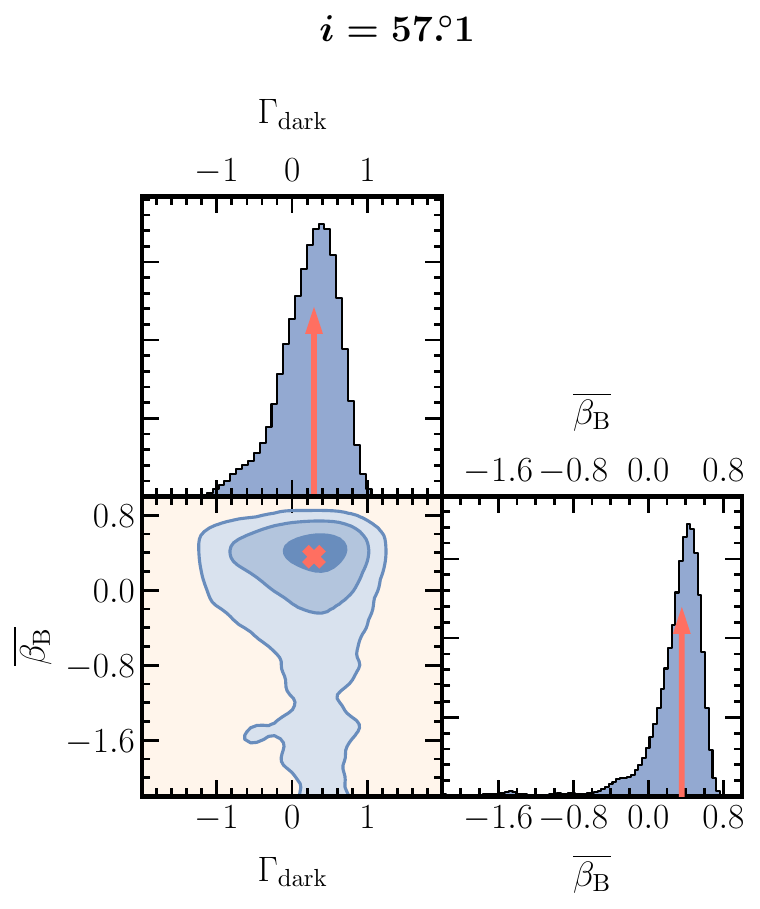}
\caption{\textit{Dark matter slope and velocity anisotropy:} Posterior probability distributions of Sculptor's dark matter density slope, $\Gamma_{\rm dark}$, averaged over the range where we have proper motion data, and the globally-averaged $\beta_{\rm B}$ velocity anisotropy (see Eq.~[\ref{eq: binney-beta}]). These results correspond to a fiducial inclination of $57\fdg1$, as argued in Section~\ref{ssec: axi-good-fit}. Curves are lightly smoothed for simple visualization purposes. The values highlighted correspond to the median estimates (arrows, and cross).}
\label{fig: gamma-dm-pdf}
\end{figure}

\begin{figure}
\centering
\includegraphics[width=\hsize]{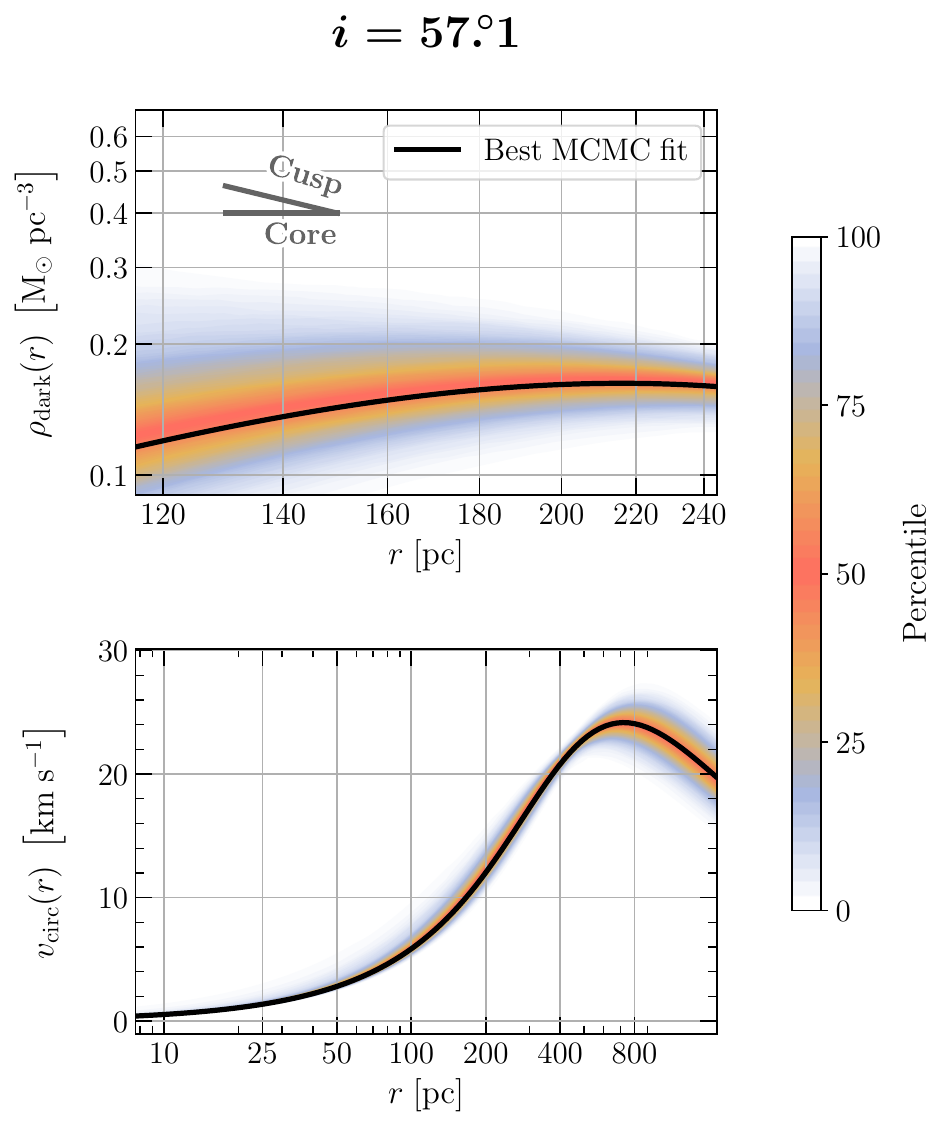}
\caption{\textit{Dark matter density and circular velocity:} The upper panel shows the 3D dark matter density profile within the radial range constrained by three-dimensional velocity data, with reference slopes for typical core and cusp profiles. The lower panel displays the corresponding circular velocity profile across the full radial extent of the data. Solid black lines indicate the median solution, while shaded colors represent percentiles of the MCMC chain. This result corresponds to a fiducial inclination of $57\fdg1$, as argued in Section~\ref{ssec: axi-good-fit}. The distance inferred for this inclination was used to convert angular to physical scales.}  
\label{fig: dm-vcirc-profile}
\end{figure}

\subsection{Data-Model Comparison} \label{ssec: axi-good-fit}

While having in mind the degeneracies in our dataset and the inclination-averaged estimates summarized in the last row of Table~\ref{tab: mass-modeling-axi}, adopting a Jeans model tied to a fiducial inclination offers a clearer benchmark for comparing data and interpreting results. For this purpose, we adopt the $i = 57\fdg1$ Jeans model as our fiducial setting, as it corresponds both to the best \scf\ match under its distribution function of type 1 and approximates the median inclination of the posterior shown in Figure~\ref{fig: inc-dist}.

Figure~\ref{fig: jeans-radial-profiles} illustrates the goodness-of-fit of the radial kinematic profiles from our Jeans models. Black points denote binned data: LOS kinematics on the left -- annularly-averaged velocity dispersion (top) and major-axis rotation (bottom); and PM kinematics on the right -- velocity dispersions in the radial (top) and tangential (bottom) directions. The solid black lines show the median model, while shaded regions represent MCMC percentiles, using the same fiducial inclination of $57\fdg1$ to convert PMs to physical velocities. Model curves are interpolated over projected radius $R$ for display, although they also vary with projected angle. The model provides a satisfactory overall agreement, consistent with fits at other inclinations.

\subsection{Inferred Quantities} \label{ssec: overall-results}

Degeneracies between inclination and parameters such as the DM inner slope and the velocity anisotropy profile render the inclination-averaged estimates of these quantities inconclusive, spanning a range of scenarios consistent with various cosmological models and DM candidates. However, incorporating higher-order moments, as outlined in Section~\ref{sssec: high-order}, allows us to favor more flattened or less inclined configurations in the Jeans modeling. In this context, confidence in shallower DM slopes becomes notable: for our fiducial inclination of $i = 57\fdg1$, we rule out $\Gamma_{\rm dark} \leq -1$ at 99.8\% confidence, while the inferred velocity anisotropy is mildly radial but still statistically consistent with isotropy. These trends are clearly reflected in the corner plots shown in  Figure~\ref{fig: gamma-dm-pdf}. Respectively, the confidence in such general patterns decreases or increases for more or less inclined systems.

The same trend is illustrated in the upper panel of Figure~\ref{fig: dm-vcirc-profile}. Over the radial range spanned by the 3D-velocity data (i.e. the limits of the respective $x$ axis) for the $i = 57\fdg1$ model, the DM density profile clearly favors a shallow inner slope. As noted earlier, edge-on configurations yield more cuspy profiles, though they are disfavored due to the higher LOS velocity kurtosis in the data. The same figure also presents Sculptor’s circular velocity profile for the fiducial inclination. Here, the velocity peaks at $\mathrm{max}(v_{\rm circ}) = 24.18^{+1.29}_{-1.19}~\kms$, followed by a decline within the data extent -- a feature not captured for Draco in Paper~I, and in good agreement with the similarly non-spherical analysis of \citet[][figure 9]{Hayashi+20}.

The DM mass enclosed within the data extent for the inclination-averaged fit -- similar to values at other inclinations -- is $M_{\rm dark}^{R_{\rm max}} = 1.39^{+ 0.68}_{- 0.58} \times 10^{8}~\msun$, with an extrapolated total mass of $M_{\rm dark}^{\rm total} = 1.58^{+ 1.01}_{- 0.62} \times 10^{8}~\msun$. This is far below the $\sim 10^{10}~\msun$ inferred by \citet*{Errani+18} for cored profiles, together with their $r_{\rm max} \sim 10$~kpc and $v_{\rm max} \sim 65~\kms$. The difference is likely explained as follows: their method matches $M(< 1.8 \, R_{\rm h})$ -- a quantity consistent with ours\footnote{\cite{Errani+18} predict $M(< 1.8 \, R_{\rm h}) = 3.2 \pm 0.7 \times 10^{7}~\msun$, while our inclination averaged measurement is $M(< 1.8 \, R_{\rm h}) = 5 \pm 2 \times 10^{7}~\msun$.} -- to subhalos in the Aquarius simulations \citep{Springel+08}, but this does not guarantee agreement in the extrapolated outer halo. Because their approach links the halo outskirts directly to the mass within $1.8 \, R_{\rm h}$, cusp and core models can produce very different $M_{\rm total}$, $r_{\rm max}$, and $v_{\rm max}$. 
Our Jeans-based fits, in contrast, allow the outer profile to vary independently of correlations inherent to cosmological simulations, although our choice of a steeper outer slope\footnote{$\beta = 5$ in the generalized $\alpha \beta \gamma$ model.} naturally yields lower total masses, even when matching at smaller radii. If Sculptor’s halo was once more massive and lost substantial DM to tidal interactions (see Section~\ref{ssec: tides}), this steep slope remains physically justified \citep{Penarrubia+10}.

Beyond inclination-sensitive parameters, some constraints remain tight despite their correlations with inclination. This is the case for our inclination-averaged distance estimate, $D = 85.83^{+ 2.02}_{- 1.91}$~kpc. This measurement is valuable in its own right, as dynamical distances provide an independent benchmark for stellar evolution models and parallax-based methods. The following sections assess the robustness of these results and their broader implications for cosmology and equilibrium-based mass modeling.

\section{Robustness of Modeling Results} \label{sec: robust}

\subsection{The Effect of Tides} \label{ssec: tides}

Perhaps the most fundamental assumption of our model is that the Sculptor dSph is in dynamical equilibrium. However, this galaxy orbits a significantly more massive system (i.e. the Milky Way) and is therefore subject to tidal interactions that could perturb its equilibrium state. This possibility has been raised in previous studies to account for the inflated velocity dispersion profiles observed in dSphs, which appear larger than expected from their baryonic content alone \citep[][]{Klessen&Kroupa&Kroupa98,Hammer+18,Wang+24}. Such arguments have even been used to question the necessity of DM in these systems, although the existence of satellites exhibiting high velocity dispersions while orbiting at relatively large pericentric distances \citep[e.g.][]{Li+21,Battaglia+22,Pace+22} remains a strong counterargument to this interpretation.

In our case, the total dark mass inferred for Sculptor is $1.58\times10^{8}~\msun$, nearly an order of magnitude lower than the $4.85\times10^{9}~\msun$ predicted by $\Lambda$CDM scaling relations for low-mass dwarfs \citep{Read+17}. This discrepancy could be somewhat accounted if a large fraction of Sculptor’s DM halo has been stripped by the Milky Way’s tidal field over its lifetime. Such a scenario is consistent with orbital reconstructions \citep[e.g.,][]{Sohn+17}, which indicate several close pericentric passages around the Milky Way, that would have favored substantial mass loss.

Sculptor’s dynamical stability under such conditions was examined by \cite{Iorio+19}, who ran simulations to evaluate how tidal interactions might affect the system and the validity of equilibrium-based mass estimates. Even under extreme orbital scenarios permitted by current observed data, they found little evidence for significant tidal disturbance in the stellar distribution. While the DM halo may undergo stripping, the stars themselves remain largely unaffected, showing only gradual changes over time. In the last 2~Gyr, they found these variations to be minimal, well within a few percent. Such findings, alongside similar conclusions from more recent studies \citep[e.g.,][]{Tchiorniy&Genina25}, support the use of equilibrium-based models for estimating Sculptor's inner mass profile.

\subsection{Tests with mock data} \label{ssec: mock}

\begin{figure}
\centering
\includegraphics[width=\hsize]{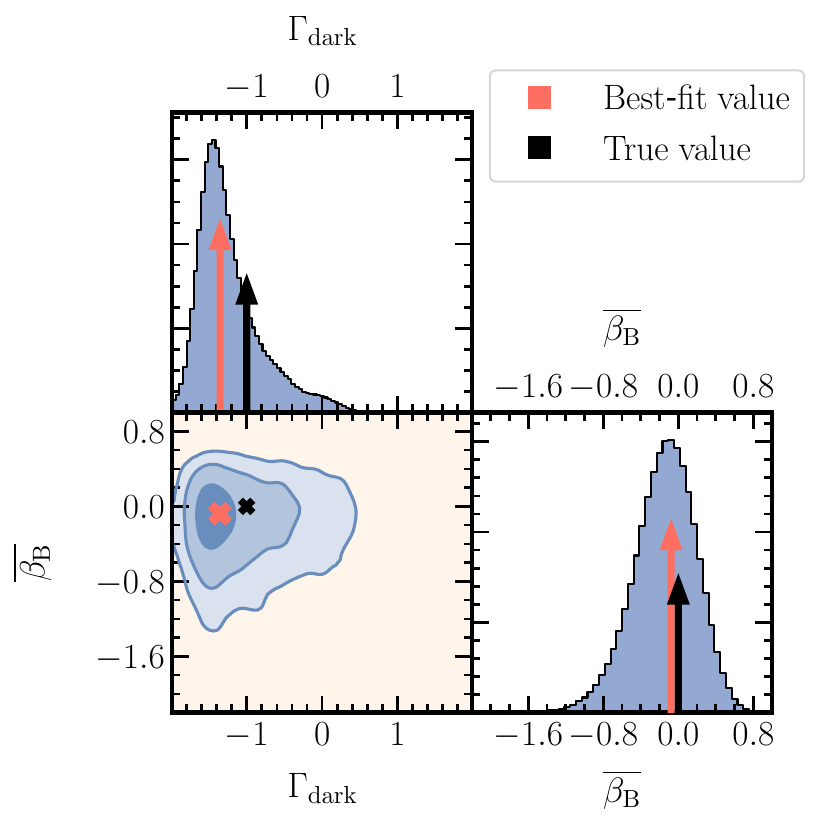}
\caption{\textit{Mock test \#1:} Corner plot of the Jeans fit to a mock dataset generated with \agama, following the overall assumptions of our modeling, for a selection of most relevant parameters: the asymptotic inner slope of the DM profile over the range where 3D velocities are available, and the average velocity anisotropy $\overline{\beta_{\rm B}}$ (Eq.~\ref{eq: binney-beta}). The black arrows and crosses (smaller) represent the true values used to construct the mock, while the larger red markers correspond to the best-fit values from our analysis. The blue posterior distributions are also shown. The close agreement between the estimates and true values indicates that our data handling -- including the removal of streaming motions through local systematic error corrections -- does not introduce significant biases, assuming the validity of our parametrization choices.}  
\label{fig: mock-correct-assumptions}
\end{figure}

\begin{figure}
\centering
\includegraphics[width=\hsize]{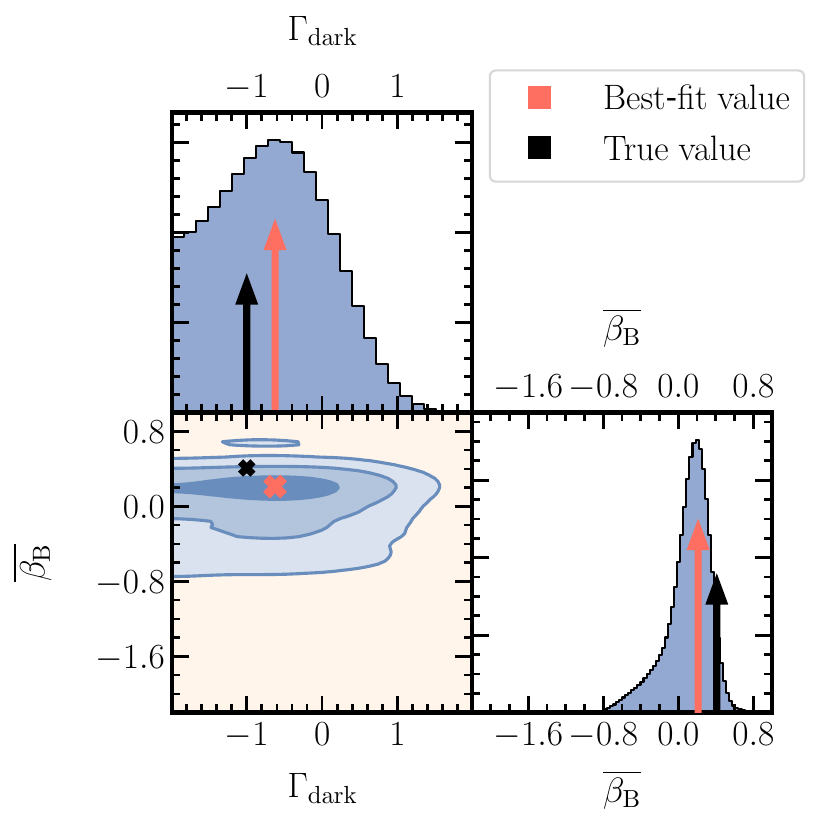}
\caption{\textit{Mock test \#2:} Same as Figure~\ref{fig: mock-correct-assumptions}, but using a mock dataset with two distinct kinematic populations (i.e., differing in both velocity anisotropy and density distribution). The average velocity anisotropy $\overline{\beta_{\rm B}}$ (Eq.~\ref{eq: binney-beta}) remains relatively well constrained thanks to the 3D velocity data, while the $\Gamma_{\rm dark}$ parameter (Eq.~\ref{eq: gamma-practicle}) shows a mild loss in fitting performance, though still consistent within 1-$\sigma$ uncertainties.
}  
\label{fig: mock-wrong-assumptions}
\end{figure}

As a final robustness check, we apply our modeling strategies to mock datasets to evaluate both their performance and the influence of key assumptions. We focus on two primary tests:
First, we simulate a system that adheres to the assumptions of our modeling framework, and examine whether elements such as the use of Gaussian likelihoods for projected velocities or our data-cleaning procedures -- particularly the local PM corrections applied to mitigate systematics -- introduce significant biases in the inferred parameters.
Second, we assess the impact of multiple kinematic populations within Sculptor, despite our modeling being based on a single-component assumption.

To conduct these tests, we employ the distribution function-based code \agama\ \citep{Vasiliev19agama} to generate spherical\footnote{At present, \agama\ does not support fully axisymmetric distributions with a given anisotropy and density profile, as would be ideal for comparison with the observed data.} stellar systems in equilibrium, given a potential and mass distribution. Each mock setup is described below, while the general procedure used to align the mocks with the observed data is detailed in Appendix~\ref{app: mocks}.

\subsubsection{Data and Likelihood Handling} \label{sssec: mocks-lcorr-lnlik}

For the first test, we construct a rotating mock system consisting of an isotropic Plummer stellar component with the same total mass and scale radius as Sculptor (see Appendix~\ref{app: mocks} for details). The DM halo is modeled using a cuspy generalized Plummer profile (i.e., $\gamma = -1$), with the total mass and scale radius set according to the expectations from $\Lambda$CDM cosmology, as outlined in Section~\ref{sssec: priors-jampy}.

By comparing our fits to this simulated data to the true input values, we can assess the impact of our local PM corrections to eliminate \hst\ systematic effects, as well as evaluate the performance of our fitting strategy. Figure~\ref{fig: mock-correct-assumptions} shows this comparison for a selection of the most relevant fit parameters: the slope of the DM profile over the region where 3D velocities are available, $\Gamma_{\rm dark}$, and the globally-averaged $\overline{\beta_{\rm B}}$ (cf. Eq.~\ref{eq: binney-beta}). The close agreement between the smaller black markers (true values) and the larger red markers (estimated values) highlights the proper handling of the data and demonstrates satisfactory convergence of our fit, assuming the parametrizations are correct. Additionally, the posterior distributions in blue exhibit similar widths to those obtained from fitting the true data, which further supports the reliability of our error bars presented in Table~\ref{tab: mass-modeling-axi}.

\subsubsection{Distinct kinematic populations} \label{sssec: mocks-chemo-kinematic}

One of the key outcomes of spectroscopic campaigns targeting dSphs, with Sculptor frequently serving as a benchmark system, has been the identification of at least two distinct chemo-dynamical populations \citep[e.g.][]{Tolstoy+04}. These subpopulations have proven instrumental in mitigating degeneracies inherent to mass modeling based solely on LOS velocity measurements \citep[e.g.][]{Walker&Penarrubia&Penarrubia11}. More recently, expanded observational efforts \citep{Tolstoy+23,Barbosa+25} have substantially advanced the characterization of Sculptor’s internal kinematics, yielding improved constraints on the spatial distributions, membership probabilities, and LOS velocity dispersion profiles of its distinct stellar populations \citep[e.g.][]{Arroyo-Polonio+24}.

This naturally raises the question of whether modeling the system as a single dynamical component introduces bias -- particularly in the estimation of key parameters such as $\Gamma_{\rm dark}$ and the mean $\beta_{\rm B}$ for the data. To assess the potential impact of multiple stellar populations, we simulate a composite system comprising metal-poor and metal-rich components. These components are designed to approximate the spatial and kinematic structure inferred by \cite{Arroyo-Polonio+25} for Sculptor, as illustrated in their figure~1, but without rotation, as not to mix different test diagnostics.
To implement this, we extracted the relevant data from that figure using online plot digitization tools,\footnote{Specifically, \url{https://web.eecs.utk.edu/~dcostine/personal/PowerDeviceLib/DigiTest/index.html}.} and fitted each population: we used a Plummer profile for the surface density and constant velocity anisotropy model for their inferred $\beta_{\rm B}$.\footnote{As discussed in Section~\ref{sssec: priors-jampy}, separate tests showed that assuming a radially varying anisotropy did not significantly alter the results presented in Table~\ref{tab: mass-modeling-axi}, thereby justifying the simpler assumption of a constant value per population. Moreover, the large $\beta_{\rm B}$ uncertainties in figure~1 of \cite{Arroyo-Polonio+25} are statistically consistent with this simplified treatment.} We then assigned tracer counts based on the normalization of each fitted density profile.

Figure~\ref{fig: mock-wrong-assumptions} presents the results of our fits to this chemo-dynamically distinct mock dataset, focusing on the same key parameters evaluated in the previous test. Overall, we find that the estimates remain consistent within 1-$\sigma$ uncertainties, albeit with slightly reduced performance compared to the single-population case. Notably, the posterior distribution of the $\Gamma_{\rm dark}$ parameter exhibits a different shape and central value, suggesting that neglecting multiple populations may slightly inflate the formal uncertainties. This behavior implies a potential bias in precision and accuracy. In summary, while the simplified single-population modeling does not introduce substantial bias in the recovery of the DM density profile, it is not entirely neutral and may slightly favor shallower inner slopes relative to Sculptor’s true underlying profile. 
Meanwhile, the average velocity anisotropy remains relatively well constrained, benefiting from the full 3D velocity information.

\section{Discussion} \label{sec: discussion}

\subsection{Implications for mass modeling} \label{ssec: disc-mass}

Nearby dSphs provide a rare opportunity to test small-scale predictions of cosmological models using resolved stellar kinematics. This has driven extensive LOS velocity surveys with ground-based telescopes \citep[e.g.,][]{Tolstoy+04,Walker+07,Gilmore+22}, aimed at constraining the underlying DM distributions. Many mass-modeling studies, however, have relied on simplifying assumptions: most notably, spherical symmetry \citep[e.g.,][]{Battaglia+08,Walker&Penarrubia&Penarrubia11,Read+18,Arroyo-Polonio+25}, despite clear observational evidence that dSphs are flattened systems \citep[e.g.,][]{Munoz+18}. Even with growing access to more complex PM data \citep[e.g.,][]{Massari+17,Massari+20}, few works have incorporated intrinsic geometry into their models (see \citealt{Hayashi+20} and \citealt{Pace+20}).

Our results, consistent with Paper~I, demonstrate that accounting for dSph flattening significantly alters the inferred DM density profile and stellar orbital structure. While Paper~I emphasized how the projected velocity anisotropy varies with position angle, potentially biasing DM slope estimates under spherical symmetry,\footnote{In spherical models, observables are assumed to be identical at a given radius regardless of position angle.} our current analysis highlights a strong dependence on the galaxy's inclination. This introduces a key degeneracy overlooked by spherical, non-rotating models. As shown in Table~\ref{tab: mass-modeling-axi}, several modeled parameters shift with inclination, in some cases beyond formal uncertainties.

These findings support the case for more flexible modeling frameworks, even those  possibly including non-spherical or triaxial DM halos and baryonic content. Yet, such approaches remain difficult in practice due to the lack of independent constraints on 3D shapes for individual dSphs. Axisymmetric models offer a pragmatic middle ground, requiring only projected flattening and inclination, with the former already well measured. Independent inclination constraints may soon be accessible through future \gaia\ data releases, which will deliver higher precision maps of mean PM, hopefully with higher signal than \gaia's systematic uncertainties. Combining these with LOS velocity data via established formalisms \citep[e.g.,][]{Evans&deZeeuw94,Bianchini+18_rotation_gc} may finally enable robust, model-independent estimates of dSph inclinations.

Another important complexity -- often not explicitly modeled, including in this study -- is the presence of multiple chemically-distinct stellar populations within dSphs. Many systems are known to host at least two such populations, typically metal-poor and metal-rich, which exhibit different LOS velocity dispersions \citep[e.g.,][]{Pace+20,Arroyo-Polonio+24}. Whether these populations require separate anisotropy prescriptions in Jeans-based modeling remains an open question, with only a limited number of systems studied in detail \citep[e.g.,][]{Arroyo-Polonio+25}. In the specific case of Sculptor, our mock data tests suggest that treating its stars as a single dynamical component does not lead to significant bias in the recovered DM density profile, though it may introduce a modest loss of accuracy within the formal 1-$\sigma$ uncertainty range.\footnote{Besides, the locations of the 16th and 84th percentiles may change, while their relative distance to each other seems somewhat the same as the one from the single population test.}

This underscores the advantages of better associating PMs with distinct chemical populations to fully exploit 3D kinematics. Progress will depend on joint astrometric and spectroscopic observations, especially at faint magnitudes where most dSph stars lie and where facilities like \hst, \jwst, \textit{Euclid} \citep[e.g.][]{Libralato+24} and the upcoming \textit{Nancy Grace Roman Space Telescope} will be crucial.

\subsection{Implications for the nature of dark matter} \label{ssec: disc-cosmo}

As discussed in Sections~\ref{ssec: axi-inc} and \ref{ssec: overall-results}, significant degeneracies arise due to Sculptor’s unknown inclination, preventing firm constraints on the nature of DM across the full range of tested configurations. These limitations stem from both the data, such as the lack of odd PM moments by design, and our modeling approach, which in the Jeans formalism does not account for velocity moments beyond second order. However, as shown in Section~\ref{sssec: high-order} and in Section~5.4 of Paper~I, higher-order moments of the LOS velocities provide additional information on velocity anisotropy and the galaxy’s shape, helping to alleviate these degeneracies \citep[e.g.][]{Read&Steger17,Read+18}. While our analysis does not yield a precise estimate of Sculptor’s inclination, comparisons with \scf\ models from \citet{deBruijne+96} suggest that Sculptor likely favors a more flattened and less inclined configuration, deviating from an edge-on geometry.

At low inclination, we find that the DM slope within the PM-covered region is constrained toward shallower values, deviating from the cuspy profiles predicted by $\Lambda$CDM DM-only simulations \citep{Navarro+97}. This result supports previous findings that favor core-like DM profiles in dSphs, based on stellar kinematics \citep[e.g.][]{Battaglia+08,Walker&Penarrubia&Penarrubia11,Breddels+13,Zhu+16,Amorisco&Evans12,Brownsberger&Randall&Randall21,Arroyo-Polonio+25} and HI rotation curves \citep[e.g.][]{Moore94,Flores&Primack94,Burkert95}. Within $\Lambda$CDM, the most common explanation involves the influence of baryonic processes, which can transform central cusps into cores by redistributing mass and energy via supernova feedback \citep{Read&Gilmore05,Pontzen&Governato12,Brooks&Zolotov&Zolotov14} or bursts of star formation \citep{Read+18}. Indeed, \citet{Fitts+17} identify a threshold around $M_{\star} \approx 2 \times 10^{6}~\msun$, above which these effects become efficient.

The comparison of total stellar masses between Draco ($4.7 \times 10^{5}~\msun$, \citealt{Martin+08}, Paper~I), where we found a clear preference for a cusp, and Sculptor ($8.08 \times 10^{6}~\msun$, \citealt{deBoer+12}, this work), where our analyses seem to favor a core, qualitatively supports this picture and presents no direct challenge to $\Lambda$CDM. Nonetheless, underlying quantitative details remain uncertain. For instance, is there a predictable analytical or empirical link between the degree of DM slope flattening (i.e., the value of $\gamma_{\rm dark}$) and a galaxy’s baryonic content? How does the host halo mass, fixed at $M_{\rm halo} \approx 10^{10}~\msun$ in \citet{Fitts+17}, modulate this process? And how long would it take for a system like Sculptor to regenerate a cuspy density profile from the apparent present-day core? \citep[e.g.][]{Laporte&Penarrubia15} 
These questions point to the need for more nuanced exploration of stellar feedback in cosmological simulations to reproduce the configurations favored by our fits. 
In parallel, our results serve as valuable observational constraints for testing both $\Lambda$CDM and alternative models in which core formation arises naturally, such as self-interacting DM (SIDM, e.g. \citealt*{Nadler+23}) with larger cross sections or Fuzzy DM with specific particle mass ranges \citep{Luu+25}.

\section{Conclusions} \label{sec: conclusion}

We present a new proper motion catalog for the Sculptor dwarf spheroidal galaxy, comprising 119 stars with proper motion uncertainties smaller than the galaxy’s intrinsic velocity dispersion (i.e. $\sim 10~\kms$) -- making it the most precise proper motion dataset compiled for this system to date. We also combined line-of-sight velocities from multiple literature sources, resulting in a sample of 1,760 stars with similarly low velocity uncertainties. Using this state-of-the-art public dataset,\footnote{We will make the data public upon paper acceptance.} we conducted axisymmetric Jeans-based mass and anisotropy modeling, complemented by robustness tests involving higher-order velocity moments and mock datasets. Below, we summarize the main conclusions of this work.

\begin{itemize}
    \item We present updated estimates for Sculptor’s center, ellipticity, position angle, and scale radius, based on fits to stellar counts from \gaia~EDR3 (Table~\ref{tab: overview}).
    \item We confirm mild oblate rotation, with major-axis amplitudes reaching up to $\sim2~\kms$ beyond $20\farcm0$ from the galaxy center, consistent with recent line-of-sight based findings \citep{Arroyo-Polonio+24}.
    \item We demonstrate that unresolved binaries have a negligible effect on the line-of-sight velocity dispersion and thus do not significantly impact the dynamical modeling.
    \item From proper motions measured in our \hst\ fields along the projected minor axis, we find average observed velocity dispersion ratios of $\left< \sigma_{\rm POSt} \right> / \left< \sigma_{\rm POSr} \right> = 1.19 \pm 0.19$ and $\left< \sigma_{\rm LOS} \right> / \left< \sigma_{\rm POS} \right> = 0.93 \pm 0.08$, where $\sigma_{\rm POS}$ is the average dispersion across both proper motion directions. The first ratio is independent of distance, while the second scales inversely with it.\footnote{We assumed here $D = 85.83$~kpc.}
    \item We perform tests with mock datasets constructed using \agama\ to confirm that our current treatment of \hst\ proper motion systematics does not introduce significant biases in the fitted parameters of our Jeans modeling. Additionally, for the specific case of Sculptor, we show that modeling its stellar component as a single dynamical population does not significantly bias the recovered DM density profile, though it may introduce a modest reduction in precision and accuracy within the formal 1-$\sigma$ uncertainty range.
    \item Our non-spherical modeling reveals a significant degeneracy tied to the unknown inclination of the galaxy -- an effect missed under spherical symmetry assumptions. This degeneracy allows acceptable fits across a wide range of dark matter profiles, from cuspy to cored and even centrally-decreasing density configurations. While we do not directly constrain the inclination, higher-order line-of-sight velocity moments provide useful limits, and comparisons with \scf\ models from \citet{deBruijne+96} favor highly flattened (i.e., less inclined) configurations.
    \item For a representative low inclination of $i = 57\fdg1$, near the median of the inferred inclination PDF (Figure~\ref{fig: inc-dist}), we find, alongside well-constrained radial velocity anisotropy, a dark matter density slope of $\Gamma_{\rm dark} = 0.29^{+ 0.31}_{- 0.41}$ within the radial extent of our 3D velocity data, ruling out $\Gamma_{\rm dark} \leq -1$ at 99.8\% confidence.\footnote{An asymptotic dark matter density slope of $-1$ represents the classic inference from dark matter only $\Lambda$CDM simulations \citep{Navarro+97}} This confidence level increases with lower inclinations and decreases for close to edge-on configurations. At the same inclination, we infer a maximum circular velocity of $\mathrm{max}(v_{\rm circ}) = 24.18^{+1.29}_{-1.19}~\kms$.
    \item At lower inclinations, our results qualitatively agree with $\Lambda$CDM predictions in which feedback in higher-mass dwarf galaxies produces shallower dark matter profiles \citep{Fitts+17}. These models predict a transition from cusp to core at $M_{\star} \approx 2 \times 10^6~\msun$, while Sculptor’s total stellar mass is $8.08 \times 10^6~\msun$ \citep{deBoer+12}. Likewise, alternative DM models such as SIDM and Fuzzy DM also predict core formation for specific ranges of cross sections and particle masses, respectively, making our measurements a useful means of narrowing these parameter spaces.
    \item Averaged over all inclinations, we estimate an enclosed mass within the radial extent of our data of $M_{\rm dark}^{R_{\rm max}} = 1.39^{+ 0.68}_{- 0.58} \times 10^8~\msun$, and a total dark matter mass of $M_{\rm dark}^{\rm total} = 1.58^{+ 1.01}_{- 0.62} \times 10^8~\msun$. The modest increase beyond $R_{\rm max}$ reflects our choice of a steep outer halo slope ($\beta = 5$), consistent with expectations for tidally affected satellites \citep{Penarrubia+10}. We also constrain a dynamical distance of $D = 85.83^{+ 2.02}_{- 1.91}$~kpc.
\end{itemize}

Following the conclusions from the first paper in this series (Paper~I), we reaffirm the growing potential for a deeper understanding of dwarf spheroidal galaxy internal kinematics. This progress relies on assembling long time-baseline proper motion datasets with the most precise astrometric facilities, such as \hst, \jwst, and \gaia, and on the development of future missions like \textit{Roman}. Ongoing work includes observations of additional systems \citep[e.g.][]{Vitral+23-UMi} and the refinement of \hst-based measurements using \jwst\ \citep[e.g.][]{2023jwst.prop.4513V,2025hst..prop17926Bennet,2025jwst.prop.9225Bennet}.\footnote{These proposals can be retrieved in the following links: \\
\url{https://ui.adsabs.harvard.edu/abs/2023hst..prop17434V/}. \\
\url{https://ui.adsabs.harvard.edu/abs/2023jwst.prop.4513V/}. \\
\url{https://ui.adsabs.harvard.edu/abs/2025hst..prop17926B/}. \\
\url{https://ui.adsabs.harvard.edu/abs/2025jwst.prop.9225B/}. \\
} With such data, significant advances in galactic dynamics and dark matter studies are within reach.

\section{Acknowledgments}

We thank the anonymous referee for providing constructive comments that improved the quality of the paper.
Eduardo Vitral acknowledges funding from the Royal Society, under the Newton International Fellowship programme (NIF\textbackslash R1\textbackslash 241973).
Support for this work was provided by NASA through grants for programs GO-12966 and GO-16737 from the Space Telescope Science Institute (STScI), which is operated by the Association of Universities for Research in Astronomy (AURA), Inc., under NASA contract NAS5-26555, and through funding to the \jwst\ Telescope Scientist Team (PI: M. Mountain) through grant 80NSSC20K0586.
We thank Justin Read and Thomas de Boer, for sharing the algorithms used in \cite{Read+17} and \cite{deBoer+12}, respectively. \\
The Digitized Sky Surveys were produced at the Space Telescope Science Institute under U.S. Government grant NAG W-2166. The images of these surveys are based on photographic data obtained using the Oschin Schmidt Telescope on Palomar Mountain and the UK Schmidt Telescope. 
The plates were processed into the present compressed digital form with the permission of these institutions. \\
This work has made use of data from the European Space Agency (ESA) mission \gaia\ (\url{https://www.cosmos.esa.int/gaia}), processed by the \gaia\ Data Processing and Analysis Consortium (DPAC, \url{https://www.cosmos.esa.int/web/gaia/dpac/consortium}). Funding for the DPAC has been provided by national institutions, in particular the institutions participating in the \gaia\ Multilateral Agreement.

This project is part of the HSTPROMO (High-resolution Space Telescope PROper MOtion) Collaboration (\url{https://www.stsci.edu/~marel/hstpromo.html}), a set of projects aimed at improving our dynamical understanding of stars, clusters, and galaxies in the nearby Universe through measurement and interpretation of PMs from \hst, \gaia, \jwst, and other space observatories. We thank the collaboration members for the sharing of their ideas and software.

\vspace{3mm}
\textit{Data availability:}
Some of the data presented in this article were obtained from the Mikulski Archive for Space Telescopes (MAST) at the Space Telescope Science Institute. The specific observations analyzed can be accessed via \dataset[DOI: 10.17909/chzv-p290]{https://doi.org/10.17909/chzv-p290}.
See Appendix~\ref{app: online-data} for the data products of this work.

%

\vspace{3mm}
\facilities{
HST; Gaia; Magellan:Clay; VLT:Kueyen.
}


\software{
{\sc Python} \citep{VanRossum09}, 
{\sc BALRoGO} \citep{Vitral21},
\jampy\ \citep{Cappellari20},
{\sc emcee} \citep{emcee},
{\sc Scipy} \citep{Jones+01},
{\sc Numpy} \citep{vanderWalt11},
{\sc Matplotlib} \citep{Hunter07},
\agama\ \citep{Vasiliev19agama},
{\sc aplpy} \citep{aplpy2012},
\scf\ \citep{deBruijne+96},
and the Scientific color maps from \cite*{Crameri+20}.
}

\bibliography{src}{}
\bibliographystyle{aasjournal}




\appendix

\section{Mock data recipe} \label{app: mocks}

\begin{figure}
\centering
\includegraphics[width=0.85\hsize]{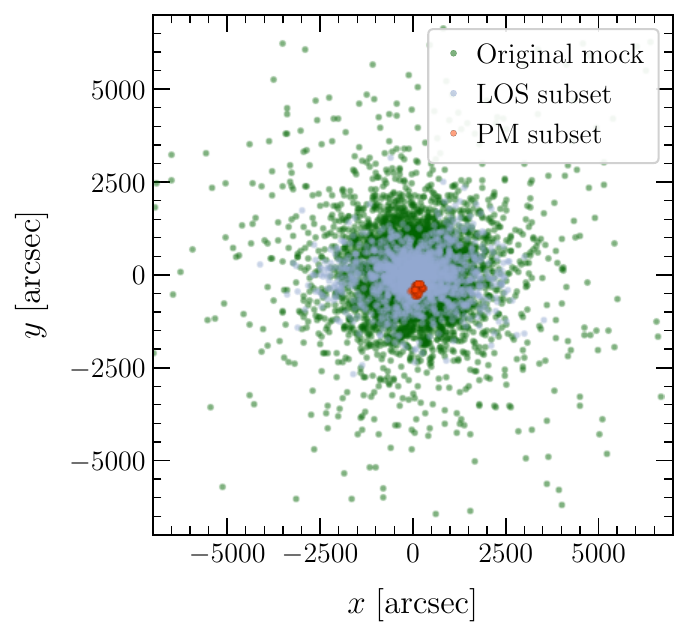}
\caption{\textit{Sky view of mock data:} The positions of the stars in our mock dataset, with $x$ and $y$ directions aligned to the galaxy's major and minor axes, are shown at three stages of the mock data construction. The original stars, sampled using the \agama\ software, are depicted in green. Stars associated with line-of-sight velocity measurements are shown in blue, while those with proper motion measurements are in red. This configuration enables us to assess the effect of spatial incompleteness on our final estimates.}
\label{fig: mock-sky}
\end{figure}

In this section, we outline the procedure used to construct realistic mock datasets from the raw output generated by the \agama\ distribution function-based software \citep{Vasiliev19agama}. This output provides phase-space information for a given kinematical population, with stars distributed according to a specified input density profile within a given potential. For context, we adopted \agama's \texttt{QuasiSpherical} distribution function,\footnote{Details of this module are available in the software documentation: \url{https://github.com/GalacticDynamics-Oxford/Agama}.} assuming constant velocity anisotropy. This was set to isotropic in mock test \#1 (cf. Section~\ref{sssec: mocks-lcorr-lnlik}), and matched to the mean anisotropies inferred for Sculptor's metal-poor and metal-rich populations from figure~1 of \cite{Arroyo-Polonio+25} in mock test \#2 (cf. Section~\ref{sssec: mocks-chemo-kinematic}). 
In mock test \#1, we also introduced a mild rotational component by flipping the velocity vectors of 20\% of the stars\footnote{This choice reflects our findings from Section~\ref{sssec: rotation}, where a similarly low rotation amplitude was observed.} with $L_{z} < 0$. To ensure that this rotation would produce observable signatures in the plane of the sky, we adopted an intermediate inclination of $i = 57\fdg1$, deviating from an edge-on configuration where PM streaming would become undetectable. This setup allowed us to reproduce a scenario broadly consistent with the inclination constraints from Section~\ref{sssec: high-order}, while still enabling a meaningful test of our artificial removal of streaming motions in the PM data.

The first step once the \agama\ outcome was ready involved converting the physical quantities of velocities and sky positions into observed, angular-based quantities, assuming a specific distance to the galaxy. We adopted the same distance value used in the priors for our Jeans modeling. Next, we applied local corrections to the PM components of the mock data, following the same procedure outlined in Section~\ref{sssec: loc-corr-implementation} and in Paper~I. This correction effectively removed any trace of transverse streaming motions from the mock data.

The next step was to select stars located within the same sky region as the real data for each subset (i.e., the LOS and PM subsets). This was accomplished by first identifying the stars within a convex hull defined by the spatial boundaries of each observed dataset. We then randomly selected the same number of stars from these regions as were present in the observations, aiming to assess the influence of spatial incompleteness on our estimates. Figure~\ref{fig: mock-sky} illustrates the spatial distribution of the stars in each of the three components: the original mock data sampled by \agama\ (shown in green), the selected LOS subset (in blue), and the PM subset (in red).

The final step was to assign realistic uncertainties to both the $v_{\rm LOS}$ and PM quantities. For the LOS velocities, we sampled errors from the 1D distribution of observed $\epsilon_{v_{\rm LOS}}$. For the PMs, we drew from the observed 2D distribution of $\epsilon_{\mu_{\alpha,*}}$ vs. $\epsilon_{\mu_{\delta}}$, which reflects both nominal measurement uncertainties and the additional scatter introduced by local systematic corrections in the real dataset. Although this procedure does not exactly replicate the real-data reduction pipeline -- where local corrections are applied after centering-related errors are already present -- it provides an effective approximation. Under the assumption of uncorrelated, Gaussian-distributed errors (which we tested in Section~\ref{sssec: loc-corr-implementation}), the statistical properties of the final error distribution remain robust to the order in which uncertainties are introduced.\footnote{This is a consequence of the fact that a convolution of two Gaussian distributions is still a Gaussian.} Moreover, drawing directly from the real observed uncertainties ensures that the mock error budget reflects realistic observational effects such as CTE degradation, which are absent in the idealized Agama mocks. We then added zero-centered Gaussian noise with these sampled standard deviations to the respective $v_{\rm LOS}$ and PM values, completing the construction of the mock dataset.

\section{Online material} \label{app: online-data}

\begin{deluxetable*}{rrrrrrrrr}
\tablecaption{Proper motion catalog.}
\label{tab: pm-catalog}
\tablewidth{750pt}
\renewcommand{\arraystretch}{1.0}
\tabcolsep=4pt
\tabletypesize{\scriptsize}
\tablehead{
\colhead{$\alpha$} & 
\colhead{$\delta$} &
\colhead{$\mu_{\alpha,*}$} &
\colhead{$\epsilon_{\mu_{\alpha,*}}$} &
\colhead{$\mu_{\delta}$} &
\colhead{$\epsilon_{\mu_{\delta}}$} &
\colhead{F775W} &
\colhead{F606W} &
\colhead{ID} \\
\colhead{[$^{\circ}$]} &
\colhead{[$^{\circ}$]} &
\colhead{[$\masyr$]} &
\colhead{[$\masyr$]} &
\colhead{[$\masyr$]} &
\colhead{[$\masyr$]} &
\colhead{[dex]} &
\colhead{[dex]} &
\colhead{} \\
\colhead{(1)} &
\colhead{(2)} & 
\colhead{(3)} & 
\colhead{(4)} & 
\colhead{(5)} & 
\colhead{(6)} & 
\colhead{(7)} & 
\colhead{(8)} &
\colhead{(9)} \\
}
\startdata
$ 14.98182$ & $-33.84174$ & $ 0.0259$ & $ 0.0189$ & $-0.0052$ & $ 0.0196$ & $-11.3723$ & $-11.4352$ & 1 \\
$ 14.99856$ & $-33.83554$ & $ 0.0238$ & $ 0.0225$ & $-0.0191$ & $ 0.0227$ & $-11.5420$ & $-11.0647$ & 2 \\
$ 15.00169$ & $-33.84224$ & $ 0.0128$ & $ 0.0189$ & $ 0.0222$ & $ 0.0155$ & $-11.3253$ & $-10.8545$ & 3 \\
\multicolumn{1}{c}{\vdots} \\
\enddata
\begin{tablenotes}
\scriptsize
\item \textsc{Notes} --  
Columns are 
\textbf{(1)} Right ascension, in degrees; 
\textbf{(2)} Declination, in degrees;
\textbf{(3)} Proper motion in right ascension, $({\rm d}\alpha / {\rm d}t)  
\cos{\delta}$, in $\masyr$;
\textbf{(4)} Uncertainty in $\mu_{\alpha,*}$, in $\masyr$;
\textbf{(5)} Proper motion in declination, $({\rm d}\delta / {\rm d}t)$, in $\masyr$;
\textbf{(6)} Uncertainty in $\mu_{\delta}$, in $\masyr$;
\textbf{(7)} Instrumental F775W magnitude, in unities of $-2.5 \log c$, where $c$ is the electron count per exposure for a source in this filter;
\textbf{(8)} Instrumental F606W magnitude, in unities of $-2.5 \log c$, where $c$ is the electron count per exposure for a source in this filter;
\textbf{(9)} Internal ID of star.
The proper motions are not absolute on the sky, but relative to motion of Sculptor, and with mean motion (i.e. first order moments) signatures removed by construction, as described in Section~\ref{sssec: loc-corr-implementation}. Hence, the proper motions are centered at ($\mu_{\alpha,*}$, $\mu_{\delta}$) = (0, 0).
A full readable version of this table will be available as online material.
\end{tablenotes}
\end{deluxetable*}

\begin{deluxetable*}{rrrrrrr}
\tablecaption{Line-of-sight velocity catalog.}
\label{tab: los-catalog}
\tablewidth{750pt}
\renewcommand{\arraystretch}{1.0}
\tabcolsep=4pt
\tabletypesize{\scriptsize}
\tablehead{
\colhead{$\alpha$} & 
\colhead{$\delta$} &
\colhead{$v_{\rm LOS}$} &
\colhead{$\epsilon_{v_{\rm LOS}}$} &
\colhead{$N_{\rm spec}$} &
\colhead{ID$_{\rm Tolstoy}$} &
\colhead{ID$_{\rm Walker}$} \\
\colhead{[$^{\circ}$]} &
\colhead{[$^{\circ}$]} &
\colhead{[$\kms$]} &
\colhead{[$\kms$]} &
\colhead{} &
\colhead{} &
\colhead{} \\
\colhead{(1)} &
\colhead{(2)} & 
\colhead{(3)} & 
\colhead{(4)} & 
\colhead{(5)} & 
\colhead{(6)} & 
\colhead{(7)} \\
}
\startdata
$ 13.66484$ & $-33.52300$ & $ 120.5981$ & $ 0.4182$ & $ 2$ & scl011\_09\_ & \multicolumn{1}{c}{--} \\
$ 13.87452$ & $-33.76831$ & $ 119.8343$ & $ 0.9640$ & $ 3$ & scl\_11\_040 & \multicolumn{1}{c}{--} \\
$ 13.93013$ & $-34.02570$ & $ 123.5945$ & $ 0.4500$ & $ 1$ & scl020-12\_ & \multicolumn{1}{c}{--} \\
\multicolumn{1}{c}{\vdots} \\
\enddata
\begin{tablenotes}
\scriptsize
\item \textsc{Notes} --  
Columns are 
\textbf{(1)} Right ascension, in degrees; 
\textbf{(2)} Declination, in degrees;
\textbf{(3)} Line-of-sight velocity, in $\kms$;
\textbf{(4)} Uncertainty in line-of-sight velocity, in $\kms$;
\textbf{(5)} Number of spectra used to construct this measurement (for matches between two catalogs, this is the sum of spectra used in each entry);
\textbf{(6)} Star ID in \protect\cite{Tolstoy+23} catalog;
\textbf{(7)} Star ID in \protect\cite{Walker+23} catalog.
\end{tablenotes}
\end{deluxetable*}

In this section, we briefly describe the kinematic datasets used in this study, which are provided as online material. These include a proper-motion table (Table~\ref{tab: pm-catalog}) derived from the \hst\ observations presented here, and a line-of-sight velocity table (Table~\ref{tab: los-catalog}) assembled by cross-matching the original catalogs of \citet{Tolstoy+23} and \citet{Walker+23}. For further details on how the data was assembled, see Sections~\ref{ssec: los} and \ref{ssec: pms}.

\end{document}